\documentclass[12pt]{article}
\usepackage{amssymb}

\usepackage{mcite}
\usepackage{makeidx}
\usepackage{jeep}
\usepackage{citesort}
\usepackage{graphics}
\usepackage[dvips]{graphicx}

\input tcilatex
\makeindex

\begin{document}

%
\date{\ }
\title{COMPETING STYLES OF STATISTICAL MECHANICS:
II. Comparison of Theory and Experiment and Further Illustrations}

\author{ \'Aurea R. Vasconcellos, J. Galv\~{a}o Ramos, and Roberto Luzzi \\
\ \\
{\small \emph{Instituto de F\'{\i}sica `Gleb Wataghin',}}\\
{\small \emph{Universidade Estadual de Campinas, Unicamp}}\\
{\small \emph{13083-970 Campinas, S\~ao Paulo,} Brazil}}
\maketitle
%

\begin{quotation}
%
%
\thispagestyle{empty}
\noindent

In the present follow-up article of a previous one [1] we illustrate the use of the
Unconventional Statistical
Mechanics described and discussed in the latter. This is done via the analysis,
resorting to Renyi's approach,
of experimental results in the case of so-called "anomalous" luminescence in
nanometric
quantum wells in semiconductor heterostructures, and the so-called "anomalous"
cyclic
voltammetry in fractal-like electrodes in microbatteries. Also a
purely theoretical analysis is
done in the cases of an ideal gas and of radiation comparing the conventional and
unconventional
approaches. In all of these situations it is discussed which is the
failure to satisfy Fisher's Criteria of Efficiency and/or Sufficiency thus requiring to resort
to the unconventional approach, and
what determines the value of the infoentropic index in each case, and its dependence
on the system characteristics.
Moreover, on the basis of the
results we obtain, it is conjectured that the infoentropic index may satisfy what we
call a law defining a "path to sufficiency".

%
\end{quotation}

\newpage

\noindent

\section{INTRODUCTION}

In a preceding article \cite{LVR02}\ (heretofore referred to as \textbf{I})
has been presented the construction of a so-called \textit{Unconventional
Statistical Mechanics} (USM), that is to say auxiliary forms for use instead
of the conventional one -- the latter based on the quite general and well
established formalism of Boltzmann and Gibbs -- when, as there noticed, the
researcher is unable to satisfy Fisher's Criteria of Efficiency and/or
Sufficiency in Statistics, namely, a failure in the characterization of the
system and its dynamics in what is relevant for the problem in hands. In the
present paper we apply the theory to some particular cases, namely two
experimental situations involving systems with fractal-like structures, and
an analysis of ideal gases.

We consider in next section measurements of \ ``anomalous'' luminescence in
nanometric quantum wells in semiconductor heterostructures. In section
\textbf{3} we describe the case of experiments of cyclic voltammetry in
microbatteries with thin-film fractal electrodes, where it is involved the
so-called\ ``anomalous'' diffusion of charges. In section \textbf{4} is
presented a study of ideal gases comparing the case when a proper
characterization is used and the one when as incomplete characterization is
used. In all cases we resort to the use of Renyi statistics \cite{Ren70,JA02}
described in \textbf{I}; as expected the infoentropic index $\alpha $ can
only be $1$ in the first situation (the criteria of efficiency and
sufficiency are satisfied), but different from $1$ in the incomplete
description when the criteria of efficiemcy and/or sufficiency are
faltering. As already noticed in \textbf{I}, the infoentropic index (when
different from $1$) depends on several characteristics of the system and its
dynamics, viz. the fractal topography, the type and size of the system's
geometry, its equilibrium or nonequilibrium thermodynamic state, the
experimental protocol, and so on. Furthermore, as we shall see below, it
appears that the infoentropic index can be related to quantities
characterizing the system and its description, through relations which can
be considered as indicating a kind of ``path to sufficiency''.

After the presentation of the five sections with the applications, in a last
section we add some additional comments and a summary of the results.

\section{``ANOMALOUS'' LUMINESCENCE}

There exists nowadays a large interest on the question of optical properties
of quantum wells in semiconductor heterostructures, which have been
extensively investigated in the last decades as they have large relevance
for the high performance of electronic and optoelectronic devices (see for
example Ref. \cite{Sin93}). To deal with these kind of systems, because of
the constrained geometry that they present (where phenomena develop in
nanometer scales) the researcher has to face difficulties with the
theoretical analysis. A most relevant question to be dealt with is the one
related to the interface roughness, usually having a kind of fractal-like
structure, that is, it is present a spatially varying confinement,\ which
leads to energies and wavefunction depending on boundary conditions which
need account for spatial correlations. As a consequence the different
physical properties of these systems appear as, say, ``anomalous'' when the
results are compared with those that are observed in bulk materials. A
particular case is the one of photoluminescence which we briefly describe
here. The conventional treatment via the well established Boltzmann-Gibbs
formalism has its application impaired because of the spatial correlations
resulting from the spatially varying confinement (as noticed above),
relevant in the characterization of the system, on which one does not have
access to (obviously the interface roughness varies from sample to sample
and one does not have any easy possibility to determine the topography of
the interface). This is then the reason why the \textit{criterion of
sufficiency} is not satisfied in this case.

Let us consider a system of carriers (electrons and holes) produced, in the
quantum well of a heterostructure, by a laser pulse. They are out of
equilibrium and their nonequilibrium macroscopic state can be described in
terms of an informational-based statistical thermodynamics \cite{LVR1}. It
is characterized by the time evolving quantities energy and density or
alternatively by the intensive nonequilibrium variables (Lagrange
multipliers in the variational approach to statistical mechanics)
quasitemperature $T_{c}^{\ast }\left( t\right) $ and quasi-chemical
potentials $\mu _{e}\left( t\right) $ and $\mu _{h}\left( t\right) $ ($e$
for electrons and $h$ for holes) \cite{AVL92}. Electrons and holes do
recombine producing a luminescence spectrum which, we recall, is
theoretically expressed as
\begin{equation}
I\left( \omega \mid t\right) \propto \sum\limits_{n,n^{\prime },\mathbf{k}%
_{\perp }}f_{n\mathbf{k}_{\perp }}^{e}\left( t\right) \ f_{n^{\prime }%
\mathbf{k}_{\perp }}^{h}\left( t\right) \ \delta \left( \hslash \Omega
-\epsilon _{n\mathbf{k}_{\perp }}^{e}-\epsilon _{n\mathbf{k}_{\perp
}}^{h}\right) \qquad ,  \label{1}
\end{equation}
where $f^{e}$ and $f^{h}$ are the populations of electrons and of holes, $%
\hslash \Omega =\hslash \omega -E_{G},$ with $\omega $ being the frequency
of the emitted photon and $E_{G}$ the energy gap, and the $\epsilon _{n%
\mathbf{k}_{\perp }}^{e\left( h\right) }$ are the electron (hole) individual
energy levels in the quantum well (index $n$ for the discrete levels and $%
\mathbf{k}_{\perp }$ for the free movement in the $x-y$ plane).

The textbook expression for the energy levels corresponding to the use of
perfectly smooth bidimentional boundaries is given by
\begin{equation}
\epsilon _{n\mathbf{k}_{\perp }}^{e\left( h\right) }=n^{2}\frac{\pi
^{2}\hslash ^{2}}{2m_{e\left( h\right) }^{\ast }L_{QW}^{2}}+\frac{\hslash
^{2}k_{\perp }^{2}}{2m_{e\left( h\right) }^{\ast }}\qquad ,  \label{2}
\end{equation}
where $L_{QW}$ is the quantum-well width and $m_{e\left( h\right) }^{\ast }$
is the effective mass and the populations $f$ take a form that resembles
instantaneous in time Fermi-Dirac distributions \cite{LVR1,AVL92}. Using
this expression in the calculation of $I$ of Eq. (\ref{1}) and comparing it
with the experimental results \cite{Laureto1} one finds a disagreement, and
then it is used the name of ``anomalous'' luminescence for these
experimental results. This is a consequence that we are using an improper
description of the carriers' energy levels -- we are not satisfying the
\textit{criterion of sufficiency} (as discussed in \textbf{I) }--, resulting
of ignoring the roughness of the boundaries (with self-affine fractal
structure \cite{FV91}) which needs be taken into account in these
nanometric-scale geometries, and then the boundary conditions to be placed
on the wavefunctions are space dependent. Hence complicated space
correlations are to be introduced, but to which we do not have access
(information), as already noticed. Hence, this limitation on the part of the
researcher breaks the sufficiency criterion, and application of the
Boltzmann-Gibbs-Shannon-Jaynes construction is impaired. As noticed in the
preceding article, one can try to circumvent the difficulty (which, we
stress once again, resides in the limitations the researcher has to possess
a proper characterization of the system and its dynamics, and not in
Boltzmann-Gibbs statistics) introducing unconventional statistics based on
parameter-dependent structural informational entropies.

To deal with the ``anomalous'' luminescence in nanometric quantum wells in
semiconductor heterostructures, we have used the unconventional statistics
that is derived from Renyi structural entropy, which depends on a single
parameter, namely, the infoentropic index $\alpha $ (see I). Taking as basic
variables, for describing the nonequilibrium thermodynamic state of the
``hot'' carrier system, the energy $E_{c}\left( t\right) $ and electron and
hole particle number (or density), $N_{e}\left( t\right) $ and $N_{h}\left(
t\right) $ \cite{AVL92} (the thermodynamically conjugated intensive
variables are the quasitemperature $T_{c}^{\ast }\left( t\right) $ and
quasi-chemical potentials $\mu _{e}\left( t\right) $ and $\mu _{e}\left(
t\right) $), and using Renyi informational entropy, as shown in I [cf.
Eq.(I-42)], we do have for the carriers' populations, to be used in Eq. (\ref
{1}), that
\begin{equation}
\bar{f}_{n\mathbf{k}_{\perp },\alpha }^{e\left( h\right) }\left( t\right)
=\left\{ \left\{ 1+\left( \alpha -1\right) \mathcal{B}_{\alpha }^{e\left(
h\right) }\left( t\right) \left[ \epsilon _{n\mathbf{k}_{\perp }}^{e\left(
h\right) }-\mu _{\alpha }^{e\left( h\right) }\left( t\right) \right]
\right\} ^{\frac{\alpha }{\alpha -1}}\pm 1\right\} ^{-1}\ .  \label{3}
\end{equation}
Here $\mathcal{B}_{\alpha }$ and $\mu _{\alpha }$ are modified forms of the
Lagrange multipliers associated to the basic variables energy and particle
number (see I). The first, $\mathcal{B}$, plays the role of a reciprocal of
a pseudotemperature, and $\mu $ of a quasi-chemical potential (see I and
\cite{AVL92}). Using these populations in the nondegenerate limit (cf. Eq.
(53) in \textbf{I}), which depend on $\epsilon _{n\mathbf{k}_{\perp
}}^{e\left( h\right) }$, that is, the ideal single-carrier energy level of
Eq. (\ref{1}), the luminescence spectrum is given by
\[
I\left( \omega \mid t\right) \propto \left[ 1+\left( \alpha -1\right) \frac{%
m_{x}}{m_{e}^{\ast }}\mathcal{B}_{\alpha }^{e}\left( t\right) \hslash \Omega %
\right] ^{\frac{\alpha }{1-\alpha }}\left[ 1+\left( \alpha -1\right) \frac{%
m_{x}}{m_{h}^{\ast }}\mathcal{B}_{\alpha }^{h}\left( t\right) \hslash \Omega %
\right] ^{\frac{\alpha }{1-\alpha }}=
\]
\begin{equation}
=\left[ 1+\left( \alpha -1\right) \beta _{eff_{\alpha }}\left( t\right)
\hslash \Omega +\left( \alpha -1\right) ^{2}\mathcal{B}_{\alpha }^{e}\left(
t\right) \mathcal{B}_{\alpha }^{h}\left( t\right) \frac{m_{x}^{2}}{%
m_{e}^{\ast }m_{h}^{\ast }}\left( \hslash \Omega \right) ^{2}\right] ^{\frac{%
\alpha }{1-\alpha }}\ ,  \label{4}
\end{equation}
where $\beta _{eff_{\alpha }}\left( t\right) =\frac{m_{h}^{\ast }}{M}%
\mathcal{B}_{\alpha }^{e}\left( t\right) +\frac{m_{e}^{\ast }}{M}\mathcal{B}%
_{\alpha }^{h}\left( t\right) $, $m_{x}^{-1}=\left[ m_{e}^{\ast }\right]
^{-1}+\left[ m_{h}^{\ast }\right] ^{-1}$,\ and $M=m_{e}^{\ast }+m_{h}^{\ast
} $. The second contribution in the last term in Eq. (\ref{4}) is much
smaller than the first, as verified \textit{a posteriori}, and then we
approximately have that
\begin{equation}
I\left( \omega \mid t\right) \propto \left[ 1+\left( \alpha -1\right) \beta
_{eff_{\alpha }}\left( t\right) \left( \hslash \omega -E_{G}\right) \right]
^{\frac{\alpha }{1-\alpha }}\qquad .  \label{5}
\end{equation}
The experiments reported in \cite{Laureto1} are time integrated, that is,
the spectrum is given by
\begin{equation}
\mathcal{I}\left( \omega \right) =\frac{1}{\Delta t}\int\limits_{t}^{t+%
\Delta t}dt^{\prime }I\left( \omega \mid t^{\prime }\right) \qquad ,
\label{6}
\end{equation}
where $\Delta t$ is the resolution time of the spectrometer. Using Eq. (\ref
{5}) in Eq. (\ref{6}), and in the spirit of the mean-value theorem of
calculus we write
\begin{equation}
I\left( \omega \right) \propto \left[ 1+\left( \alpha -1\right) \bar{\beta}%
_{eff_{\alpha }}\left( \hslash \omega -E_{G}\right) \right] ^{\frac{\alpha }{%
1-\alpha }}\qquad ,  \label{7}
\end{equation}
introducing the mean value $\bar{\beta}_{eff_{\alpha }}$ (as an open
parameter), which we rewrite as $\left[ \bar{\beta}_{eff_{\alpha }}\right]
^{-1}=k_{B}\Theta _{\alpha }$, defining an average, over the resolution time
$\Delta t$, effective temperature of the nonequilibrium carriers, that is, a
measure of their average kinetic energy (see \cite{Net94}).

In Fig.\textbf{1} is shown the fitting of the experimental data with the
theoretical curve as obtained from Eq. (\ref{7}). It contains the results
referring to four samples having different values of the quantum well width.
The corresponding values of $\alpha $, and the kinetic temperature $\Theta
_{\alpha }$ are given in Table \textbf{I}. The information-entropic index $%
\alpha $ depends, as expected, on the dimensions of the system: as the width
of the quantum well increases the values of $\alpha $ keep increasing and
tending to $1$. This is a clear consequence that the fractal-like
granulation of the boundary surface becomes less and less relevant, for
influencing the outcome of the phenomenon, as the width of the quantum well
falls outside the nanometer scale, and is approached the situation of a
normal bulk sample. On the other hand the kinetic temperature of the
carriers is smaller with increasing quantum well width, as also expected
once the relaxation processes, mainly as a result of the interaction with
the phonon system, become more effective and the cooling down of the hot
carriers proceeds more rapidly.

Moreover, we can empirically derive what we term as a law of ``\textit{path
to sufficiency}'', namely,
\begin{equation}
\alpha \left( L\right) \simeq \frac{L+L_{1}}{L+L_{2}}\qquad ,  \label{8}
\end{equation}
where, by best fitting, $L_{1}\simeq 139\pm 17$ $L_{2}\simeq 204\pm 24$, all
values given in \AA ngstrom. We do have here that as $L$ largely increases,
the entropic index tends to $1$, when one recovers\ the expressions for the
populations in the conventional situation (see I), but as $L$ decreases $%
\alpha $ tends to a finite value $L_{1}/L_{2}$, in this case $\sim 0.7\pm
0.06.$ This indicates that the insufficiency of description when using Eq. (%
\ref{2}) in the calculations (the ideal energy levels) becomes less and less
relevant as the size of the system increases as already commented.

It is relevant to notice that there exists computer-modelled experiments, in
which a certain controlled roughness of the quantum-well boundaries is
introduced, Schr\"{o}dinger equation is solved and the corresponding energy
levels are obtained \cite{RZ98}. The conventional statistics is applied, and
the results, for this model, qualitatively agree with the experimental ones
in real systems. Furthermore, the results of the computer modelled system,
with the sufficiency criterion being satisfied, can be reproduced using the
ideal eigenvalues of Eq.(\ref{2}) but in the Renyi statistics we used
adjusting the infoentropic index $\alpha $ \cite{Laureto1}. Another
observation is that experiments in which are present quantum wells with the
same width, but obtained with careful and improved methods of growing, show
that as smoother and smoother the boundary surfaces the entropic index $%
\alpha $ increases tending to $1$, the value corresponding to the ideal
situation of perfectly smooth surfaces \cite{Laureto1}. Moreover, from the
experimental data \cite{Laureto1} it can be noticed that the linewidth $%
\Gamma $, in each of the four samples of Fig. \textbf{1}, appears to depend
on temperature $T$, for $T>20K$, through a power law, say, $\Gamma \sim
T^{\nu }$, with $\nu $ approaching $1$ (the conventional result) as the
quantum-well width increases, as expected; study of the connection of $\nu $
and the infoentropic index $\alpha $ is under way.

As a consequence, it can be noticed that through measurements of
luminescence it is possible to obtain an evaluation of the microroughness of
the samples being grown, what implies in a kind of method for quality
control. On this we may comment that the interface structural properties in
quantum wells (QW's) have been extensively investigated, as they are
extremely important for the high performance of electronic and
optoelectronic QW-based devices\cite{SB85,BGCMCPP96,FKT89}. Semiconductor
heterostructures interfaces have been investigated by means of direct or
indirect characterization techniques. Direct investigations of the
interfacial quality have been obtained, for instance, by scanning tunneling
microscopy, atomic force microscopy, and transmission electron microscopy
\cite{JKRHGR96}. However, interfaces are not easily accessible through these
direct investigation methods, so optical techniques (which indirectly probe
the interfaces) can constitute a useful approach in semiconductor-interface
characterization in heterostructures, as shown above.

This first illustration of the theory clearly evidences the already stated
fact that the infoentropic index $\alpha $ is not a universal one for a
given system, but it depends on the knowledge of the correct dynamics, the
geometry and size including the characteristics and influence of the
boundary conditions (e.g. the fractality in the present case), the
macroscopic (thermodynamic) state of the system of equilibrium or
nonequilibrium conditions, and the experimental protocol.

The case we presented consisted of experiments in time-integrated optical
spectroscopy. The phenomenon of ``anomalous'' luminescence in nanometric
quantum wells in semiconductor heterostructures, is also present in the case
of time-resolved experiments (nanosecond time resolution where the
infoentropic index and the (nonequilibrium) kinetic temperature change in
time accompanying the irreversible evolution of the system \cite{Laureto2}).
The use of USM in a completely analogous way as done above (note that Eq. (%
\ref{4}) is valid for a time-resolved situation) allows to determine the
evolution in time of the kinetic temperature $\Theta _{\alpha }\left(
t\right) $, and the \textit{infoentropic index }$\alpha \left( t\right) $%
\textit{, which is then changing in time} as it proceeds the evolution of
the irreversible processes in the nonequilibrium thermodynamic state of the
carriers, which is a work under way.

\section{``ANOMALOUS'' DIFFUSION}

As a second illustration we consider the question of behavior of fractal
electrodes in microbatteries. As a consequence of the nowadays large
interest associated to the development of microbatteries (see for example
Ref. \cite{JN94}), the study of growth, annealing, and surface morphology of
thin-film depositions used in cathodes, has acquired particular relevance.
These kind of systems are characterized by microroughneessed surface
boundaries in a geometrically constrained region (nanometric thin films),
and then fractal characteristics can be expected to greatly influence the
physical properties \cite{FV91}. In other words the dynamics or
hydrodynamics involved in the functioning of such devices can be expected to
be governed by some type or other of scaling laws . Such characteristics
have been experimentally evidenced and the scaling laws determined.
Researchers have resorted to the use of several experimental techniques:
atomic force microscopy allows to obtain the detailed topography of the
surface -- and then measuring the fractality of it --; cyclic voltammetry is
an electrochemical technique used for the study of several phenomena, also
allowing for the characterization of the fractal characteristics (scaling
laws) of the system \cite{PN89,DK01,KDFG99}. Particularly, through cyclic
voltammetry it can be put into evidence the property of the so-called
``anomalous'' diffusion in these systems, what we consider here.

The difference of chemical potential between an anode and a cathode with a
thin film (nanometric fractal surface) of, say, nickel oxides, produces a
movement of charges in the electrolyte from the former to the latter. In a
cyclic voltammetry experiment these charges circulate as a result of the
application of an electric field, $e\left( t\right) $, with particular
characteristics: It keeps increasing linearly in time as $e_{o}+vt$, where $%
v $ is a scanning velocity, during an interval, say $\Delta t$, and next
decreases with the same scanning velocity, i.e. $e_{o}+v\Delta t-vt$, until
recovering the value $e_{o}$. A current $i\left( t\right) $ is produced in
the closed circuit, which following the field $e\left( t\right) $ keeps
increasing up to a peak value $i_{p}$, and next decreases. This current is
the result of the movement of the charges that keep arriving to the thin
film fractal-like cathode.

It is found that there follows a power law relation between the peak value, $%
i_{p}$, of the current and the rate of change, $v$, of the electric field,
namely $i_{p}\sim v^{\xi }$. Such current in the fractal electrode,
generated by the application of the external potential, depends on the
charges that are brought to the interface. Such density of charges $n\left(
x,t\right) $ at the interface is the one accumulated as a result of a
diffusive motion, generated by the difference of chemical potentials between
both electrodes. Using the standard Fick's law it follows that the power
index $\xi $ must be $1/2.$ Observation shows that it departs from that
value and to account for the disparity it has been postulated an
``anomalous'' diffusion law of the type
\begin{equation}
\frac{\partial }{\partial t}n\left( \mathbf{r},t\right) -D_{\gamma }\ \nabla
^{2}n^{\gamma }\left( \mathbf{r},t\right) =0\qquad ,  \label{10}
\end{equation}
where when $\gamma =1$ is recovered the standard result. Such kind of result
can be derived using the USM of \textbf{I}, and we resort to the statistics
based on Renyi structural information-entropy, as done in the previous
section. The question now is which is the source of the \textit{lack of
sufficiency} in the well established Boltzmann-Gibbs formalism, which forces
us to resort to the unconventional approaches. The answer resides in that is
being used a quite incomplete hydrodynamic approach to describe the motion
of the fluid. Fick's law is an approximation which gives good results under
stringent conditions imposed on the movement (one being the limit of very
large wavelengths, that is, the movement is characterized by being of smooth
variation in space (near uniform), see Appendix \textbf{A}). Motion of the
particles (charges) towards the thin film in the cathode proceeds through
the microroughneessed fractal region, and then involves a description
requiring to consider intermediate to short wavelengths. Hence, for its
description one needs to remove the limitations that restrict the movement
to be purely diffusive, that is, to introduce a higher-order generalized
hydrodynamics \cite{JCVVML02,MVL98a} (see also section \textbf{4}). This
means to introduce as basic hydrodynamic variables not only the density, $%
n\left( \mathbf{r},t\right) $, but its fluxes of all order, $\mathbf{I}%
_{n}\left( \mathbf{r},t\right) $ and $I_{n}^{\left[ r\right] }\left( \mathbf{%
r},t\right) $, where $r\geq 2$ indicates the order of the flux and its
tensorial rank: see Eqs. (\ref{31}) to (\ref{34}) in Section \textbf{4}. The
motion is then determined by a complicated set of equations of evolutions of
the type \cite{LVR1,JCVVML02,LVR2}
\begin{equation}
\frac{\partial }{\partial }I_{n}^{\left[ r\right] }\left( \mathbf{r}%
,t\right) +\nabla \cdot I_{n}^{\left[ r+1\right] }\left( \mathbf{r},t\right)
=\mathcal{J}_{n}^{\left[ r\right] }\left( \mathbf{r},t\right) \qquad ,
\label{11}
\end{equation}
where $r=0$ for the density, $r=1$ for the first (vectorial) flux, or
current, and $r\geq 2$ for the all higher-order fluxes, $\mathcal{J}_{n}^{%
\left[ r\right] }$ are the collision operators, and $\nabla \cdot $ is the
operator indicating to take the divergence of the tensor. Solving this set
of coupled equations of evolution is a formidable, almost unmanageable,
task. As a rule one uses, depending on each experimental situation, a
truncation in the set of equations (i.e. they are considered from $r=0$ up
to a certain value, say $n$, of the order $r$). Furthermore, to build and
solve the set of equations with the spatially quite complicated interface
boundary conditions is practically not possible (at most can be attempted in
a computer-modelled system). Hence, introducing such truncation produces a
failure of the \textit{criterion of sufficiency} when using the
conventional, and universal, approach. Consequently, a way to circumvent the
problem is, as done in the previous section, to make calculations in terms
of unconventional statistical mechanics, what we do resorting to Renyi's
approach.

We take a truncated description introducing as basic variables the densities
of energy and particles and the first (vectorial) particle flux, the
corresponding dynamical operators being
\begin{equation}
\hat{H}_{o}=\int d^{3}r\ \hat{h}\left( \mathbf{r\mid }\Gamma \right) \qquad ,
\label{11a}
\end{equation}
with
\begin{equation}
\hat{h}\left( \mathbf{r\mid }\Gamma \right) =\sum\limits_{i=1}^{N}\frac{%
p_{i}^{2}}{2m}\ \delta \left( \mathbf{r}-\mathbf{r}_{i}\right) \qquad ,
\label{11b}
\end{equation}
and
\begin{equation}
\hat{n}\left( \mathbf{r\mid }\Gamma \right) =\sum\limits_{i=1}^{N}\delta
\left( \mathbf{r}-\mathbf{r}_{i}\right) \qquad ,  \label{11c}
\end{equation}
\begin{equation}
\mathbf{\hat{I}}_{n}\left( \mathbf{r\mid }\Gamma \right)
=\sum\limits_{i=1}^{N}\frac{\mathbf{p}_{i}}{m}\ \delta \left( \mathbf{r}-%
\mathbf{r}_{i}\right) \qquad .  \label{11d}
\end{equation}
\qquad

The unconventional statistical operator is in this case given by Eq. (I.21)
using in it the auxiliary (``instantaneous frozen'' or quasiequilibrium)
statistical operator
\[
\bar{\varrho}_{\alpha }\left( \Gamma \mid t,0\right) =\frac{1}{\bar{\eta}%
_{\alpha }\left( t\right) }\left\{ 1+\left( \alpha -1\right) \left[ \beta
\hat{H}_{o}+\int d^{3}r\ \tilde{F}_{n}\left( \mathbf{r},t\right) \hat{n}%
\left( \mathbf{r\mid }\Gamma \right) +\right. \right.
\]
\begin{equation}
\left. \left. +\int d^{3}r\ \mathbf{\tilde{F}}_{n}\left( \mathbf{r},t\right)
\cdot \mathbf{\hat{I}}_{n}\left( \mathbf{r\mid }\Gamma \right) \right]
\right\} ^{\frac{1}{1-\alpha }}\qquad ,  \label{12}
\end{equation}
where $\tilde{F}_{n}\left( \mathbf{r},t\right) $ and $\mathbf{\tilde{F}}%
\left( \mathbf{r},t\right) $ are modified Lagrange multipliers that the
variational method introduces, $\beta =1/k_{B}T$ with $T$ being the
temperature of the system, and then we are assuming that the material motion
does not affect the thermal equilibrium, and $\bar{\eta}_{\alpha }\left(
t\right) $ ensures the normalization of the distribution, $\Gamma $ is a
point in phase space and we shall designate by $d\Gamma $ the element of
volume in phase space (taken adimensional).

The equations of evolution for the basic variables, derived in the context
of the kinetic theory based on the Nonequilibrium Statistical Ensemble
Formalism (NESEF) \cite{LVR1,JCVMVL02,LVR02} are
\begin{equation}
\frac{\partial }{\partial t}n\left( \mathbf{r},t\right) =-\nabla \cdot
\mathbf{I}_{n}\left( \mathbf{r},t\right) \qquad ,  \label{13}
\end{equation}
\begin{equation}
\frac{\partial }{\partial t}\mathbf{I}_{n}\left( \mathbf{r},t\right)
=-\nabla \cdot I_{n\alpha }^{\left[ 2\right] }\left( \mathbf{r},t\right) +%
\mathbf{J}_{n\alpha }\left( \mathbf{r},t\right) \quad ,  \label{14}
\end{equation}
where the second order flux is given in Eq. (\ref{I24}) of Appendix \textbf{A%
} and $\nabla \cdot $ is the tensorial divergence operator. Equation (\ref
{13}) is the conservation equation for the density; the terms with the
presence of the divergence operator arise out of the contribution resulting
from performing the, in this classical case, Poisson bracket with the
kinetic energy operator $\hat{H}_{o}$; and
\begin{equation}
\mathbf{J}_{n\alpha }\left( \mathbf{r},t\right) =\int d\Gamma \left\{
\left\{ \mathbf{\hat{I}}_{n}\left( \mathbf{r\mid }\Gamma \right) ,\hat{H}%
^{\prime }\left( \Gamma \right) \right\} \stackrel{-}{\mathcal{D}}_{\alpha
}\left\{ \bar{\varrho}_{\alpha }\left( \Gamma \mid t,0\right) \right\}
\right\}  \label{15}
\end{equation}
is a scattering operator in the Markovian approximation accounting for the
effects of the collisions generated by the interactions with the surrounding
media via an interaction Hamiltonian $\hat{H}^{\prime }$, and where [cf. Eq.
(\ref{I27})]
\begin{equation}
\stackrel{-}{\mathcal{D}}_{\alpha }\left\{ \bar{\varrho}_{\alpha }\left(
\Gamma \mid t,0\right) \right\} =\left[ \bar{\varrho}_{\alpha }\left( \Gamma
\mid t,0\right) \right] ^{\alpha }/Tr\left\{ \left[ \bar{\varrho}_{\alpha
}\left( \Gamma \mid t,0\right) \right] ^{\alpha }\right\} \qquad ,
\label{15a}
\end{equation}
that is, the escort probability that the unconventional statistics requires
to be used in the calculation of averages values (cf. Eq. (18) in \textbf{I}%
). To solve the system of Eqs. (\ref{13}) and (\ref{14}), we need in Eq. (%
\ref{15}) to express the right-hand side in terms of the basic variables.
The scattering operator takes in general the form of the kind that is
present in the relaxation-time approach, as for example shown in Refs. \cite
{LVR2,JCVMVL02,RVL01b}, namely
\begin{equation}
\mathbf{J}_{n\alpha }\left( \mathbf{r},t\right) \equiv -\mathbf{I}_{n}\left(
\mathbf{r},t\right) /\tau _{I\alpha }\qquad ,  \label{16}
\end{equation}
where $\tau _{I\alpha }$ is the momentum relaxation time (see Ref. \cite
{VAL93}, where it is presented an analysis of diffusion in the photoinjected
plasma in semiconductors dealt with in the conventional statistical
mechanics once the sufficiency condition is verified).

Transforming Fourier in time Eq. (\ref{14}) we have, after using Eq. (\ref
{16}), that
\begin{equation}
\left( 1+i\omega \tau _{I\alpha }\right) \mathbf{I}_{n}\left( \mathbf{r}%
,\omega \right) =-\tau _{I\alpha }\ \nabla \cdot I_{n\alpha }^{\left[ 2%
\right] }\left( \mathbf{r},\omega \right) \qquad ,  \label{17}
\end{equation}
which in the limit of small frequency, meaning $\omega \tau _{I\alpha }\ll 1$%
, becomes, after back transforming to the time coordinate,
\begin{equation}
\mathbf{I}_{n}\left( \mathbf{r},t\right) =-\tau _{I\alpha }\ \nabla \cdot
I_{n\alpha }^{\left[ 2\right] }\left( \mathbf{r},t\right) \qquad ,
\label{18}
\end{equation}
and then, after using Eq. (\ref{18}), Eq. (\ref{13}) becomes
\begin{equation}
\frac{\partial }{\partial t}n\left( \mathbf{r},t\right) =\tau _{I\alpha }\
\nabla \cdot I_{n\alpha }^{\left[ 2\right] }\left( \mathbf{r},t\right)
\qquad .  \label{19}
\end{equation}

In order to close Eq. (\ref{19}) we need to express the second-order flux in
terms of the basic variables, $n$ and $\mathbf{I}_{n}$, and after some
calculus we find that (cf. Eq. (\ref{I26}$)$ in Appendix \textbf{A} where
some additional comments and considerations are presented)
\begin{equation}
\nabla \cdot I_{n\alpha }^{\left[ 2\right] }\left( \mathbf{r},t\right) =\xi
_{\alpha }\mathbf{\nabla \ }n^{\gamma _{\alpha }}\left( \mathbf{r},t\right)
\qquad ,  \label{20}
\end{equation}
where power index $\gamma _{\alpha }$ has the $\alpha $-dependent expression
\begin{equation}
\gamma _{\alpha }=\frac{5-3\alpha }{3-\alpha }\qquad ,  \label{21}
\end{equation}
where the values of $\alpha $ are restricted to the interval $1\leq \alpha <%
\frac{5}{3}$, (see Appendix \textbf{A}) and consequently it follows the
``anomalous'' diffusion equation as given by Eq. (\ref{10}), where $D=\xi
_{\alpha }\tau _{I\alpha }$, once we take $\xi _{\alpha }$ and $\tau
_{I\alpha }$ as varying slowly in space and time, neglecting then their
dependence on $\left( \mathbf{r},t\right) $.

Let us analyze this question of anomalous cyclic voltammetry in fractal
electrodes in terms of the previous results. The solution of Eq. (\ref{10}),
for movement in one dimension, say, in $x$-direction normal to the electrode
surface, is given, for $0<\gamma \leq 1$, by \cite{JCVCS01}
\begin{equation}
n\left( x,t\right) =b_{\alpha }t^{-\mu _{\alpha }}\left[ a^{2}+x^{2}t^{-2\mu
_{\alpha }}\right] ^{\frac{1}{\gamma _{\alpha }-1}}\qquad ,  \label{22}
\end{equation}
where $a$ and $b_{\alpha }$ are constants and $\mu _{\alpha }=\left( \gamma
_{\alpha }+1\right) ^{-1}=\frac{1}{4}\left( 3-\alpha \right) /\left(
2-\alpha \right) .$ Hence, taking into account that the current, as noticed,
is proportional to the arriving charges, and, once $\mu _{\alpha }>0,$%
admitting that $x^{2}t^{-2\mu _{\alpha }}\ll a^{2}$, there follows that
\begin{equation}
I\left( t\right) \approx \frac{b_{\alpha }}{\gamma _{\alpha }-1}a^{\frac{2}{%
\gamma _{\alpha }-1}}t^{-\frac{1}{\gamma _{\alpha }+1}}=\frac{1}{2}\left[
\frac{3-\alpha }{1-\alpha }\right] \ b_{\alpha }\ a^{\frac{3-\alpha }{%
1-\alpha }}t^{-\frac{1}{4}\frac{3-\alpha }{2-\alpha }}\qquad .  \label{23}
\end{equation}
Finally, taking into account that the applied field is $e\left( t\right)
=e_{o}+vt$ (where $v$ is the scanning velocity), and then $t=\left(
e-e_{o}\right) /v$, Eq. (\ref{23}) leads to the potential law
\begin{equation}
I\sim v^{\xi }\qquad ,  \label{24}
\end{equation}
which stands for the peak value in the experiment and where $\xi $ is $\mu
_{\alpha }$, that is
\begin{equation}
\xi =\left( \gamma _{\alpha }+1\right) ^{-1}=\frac{1}{4}\left[ \frac{%
3-\alpha }{2-\alpha }\right] \qquad ,  \label{25}
\end{equation}
then it has been proved the empirical law, with the power $\xi $ expressed
in terms of the infoentropic index $\alpha $; the latter is then determined
from the experimental value of $\xi $, which should be contained in the
interval $0.5\leq \xi <1$, once $1\leq \alpha <5/3$.

In Fig. \textbf{2} are shown the experimental results linking the values of
the peak current for varying values of the scanning velocity $v$, from which
can be derived the values of $\xi $ for each $v$, and then those of $\alpha $
through Eq. (\ref{25}). Hence, keeping all other characteristics of the
experiment fixed, the entropic index is dependent on the experimental
protocol, in this case on $v$. From Fig. \textbf{2} and Eq. (\ref{21}), it
can be derived (similarly to what was done in the previous section) a
\textit{``path to sufficiency''}, given in this case by
\begin{equation}
\alpha \left( v\right) \simeq \frac{v+v_{1}}{v+v_{2}}\qquad ,  \label{26}
\end{equation}
where, by best fitting, $v_{1}\simeq 235\pm 20$ and $v_{2}\simeq $ $145\pm
15 $, in $mV/s.$ As we can see, as $v$ largely increases $\alpha $ goes to $%
1 $ -- when one recovers the conventional result, that is, the standard
Fick's law, indicating that the \textit{Criterion of Sufficiency} in the
conventional Boltzmann-Gibbs formalism is satisfactorily verified -- while
for small values of $v$, $\alpha $ tends \ to the lower bound value $%
v_{1}/v_{2}$ ($\simeq 1.6\pm 0.15$ in the experiment considered above). We
can interpret this as a consequence that a large rate of change of the
electric field leads to a very rapid transit of the charge through the
roughened layer at the boundary giving no time for the movement tending to
get adapted to the morphology of the thin film boundary.

In the present type of experiments in electrochemistry we can derive the
same conclusion as the one of the previous section in the experiments in
semiconductor physics: the infoentropic index $\alpha $ is not an universal
one for a given system but depends on the correct dynamics, geometry and
size as well as the characteristics and influence of the boundary
conditions, the macroscopic (thermodynamic) state of the system, and the
experimental protocol.

Two other interesting problems belong to this question of an ``anomalous''
diffusion-advection in polymer solutions and ``anomalous'' current in
ionic-conducting glasses \cite{VRLJCV03}. In the first case, macromolecules
under flow, one faces the difficulty in the description resulting from the
self-similarity in a type of average fractal structure (what we have called
``Jackson Pollock Effect'', in view of the anology with his painting with
the dripping method, leading to fractal structures \cite{Tay02}). Without
entering into details, given elsewhere \cite{VRLJCV03}, the ``anomalous''
diffusion-advection equation has the form
\begin{equation}
\frac{\partial }{\partial t}n\left( \mathbf{r},t\right) -D_{\alpha }\nabla
^{2}n^{\gamma _{\alpha }}\left( \mathbf{r},t\right) =-\tau _{\alpha }\nabla
\cdot \nabla \cdot \left( n\left( \mathbf{r},t\right) \left[ \mathbf{v}%
\left( \mathbf{r,}t\right) \mathbf{v}\left( \mathbf{r,}t\right) \right]
\right) \qquad ,  \label{9}
\end{equation}
where $\gamma _{\alpha }$ is the one of Eq. (\ref{21}), $D_{\alpha }$ of Eq.
(\ref{10}).

In the second case, ionic-conducting glasses, the difficulty in dealing with
them consists into the presence of long-range correlations in space,
mediated by Coulomb interaction, having scaling characteristics. It leads to
the so-called Curie-von Schwaitler current
\begin{equation}
J\left( t\right) \sim t^{-\nu }\qquad ,  \label{9a}
\end{equation}
that is, following a power law in time. Again, resorting to USM (in Renyi
approach) while ignoring the scaling characteristic, we arrive at such power
law, with power $\nu $ related to the infoentropic index by the expression
\begin{equation}
\nu =\frac{3-\alpha }{\left( \alpha -1\right) ^{2}}\qquad .  \label{9b}
\end{equation}

\section{IDEAL GAS IN INSUFFICIENT DESCRIPTION}

As additional illustrations we consider, first, an ideal gas of $N$
particles in a volume $V$, but in the thermodynamic limit and we call $%
\mathbf{r}_{j}$ and $\mathbf{p}_{j}$ the position and velocity of the $j$-th
particle, with $\hat{H}\left( \Gamma \right) =\sum\limits_{j}p_{j}^{2}/2m$
being the Hamiltonian. The most general description of this system in any
circumstances is in terms of all the observables of the system, that is from
the knowledge of the single-particle dynamical function \cite{LVR1}, which
in classical mechanics is given by
\begin{equation}
\hat{n}_{1}\left( \mathbf{r},\mathbf{p}\mid \Gamma \right)
=\sum\limits_{j=1}^{N}\delta \left( \mathbf{r}-\mathbf{r}_{j}\right) \
\delta \left( \mathbf{p}-\mathbf{p}_{j}\right) \qquad .  \label{27}
\end{equation}
The statistical operator depends on it and, in NESEF (see \textbf{I}), on
the associated Lagrange multiplier which we call $\varphi _{1}\left( \mathbf{%
r},\mathbf{p};t\right) $, determined by the informational constraints
consisting in the average values
\begin{equation}
n_{1}\left( \mathbf{r},\mathbf{p};t\right) =\int d\Gamma \ \hat{n}_{1}\left(
\mathbf{r},\mathbf{p}\mid \Gamma \right) \ \bar{\varrho}\left( \Gamma \mid
t,0\right) \qquad ,  \label{28}
\end{equation}
and we do have in the conventional formalism
\begin{equation}
\bar{\varrho}\left( \mathbf{\Gamma }\mid t,0\right) =Z^{-1}\left( t\right)
\exp \left\{ -\int d^{3}r\ d^{3}p\ \varphi _{1}\left( \mathbf{r},\mathbf{p}%
;t\right) \ \hat{n}_{1}\left( \mathbf{r},\mathbf{p}\mid \Gamma \right)
\right\} \quad ,  \label{29}
\end{equation}
\ for the auxiliary (``instantaneous frozen quasiequilibrium'') statistical
operator, with the nonequilibrium statistical operator given in terms of it
by Eqs (\textbf{I}.14) to (\textbf{I}.17) in \textbf{I}.

As shown elsewhere \cite{LVR1,MVL98a} we can alternatively write the
statistical operator in the form of a generalized grand-canonical one, the
associated auxiliary operator being
\[
\bar{\varrho}_{GC}\left( \Gamma \mid t,0\right) =Z_{GC}^{-1}\left( t\right)
\exp \left\{ -\int d^{3}r\ d^{3}p%
\bigl[%
%
\beta \left( \mathbf{r},t\right) \hat{h}\left( \mathbf{r}\right) +A\left(
\mathbf{r},t\right) \hat{n}\left( \mathbf{r}\right) +\right.
\]
\[
+\mathbf{\nu }_{h}\left( \mathbf{r},t\right) \cdot \mathbf{\hat{I}}%
_{h}\left( \mathbf{r}\right) +\mathbf{\nu }_{n}\left( \mathbf{r},t\right)
\cdot \mathbf{\hat{I}}_{n}\left( \mathbf{r}\right)
\]
\begin{equation}
\left. +\sum\limits_{r\geq 2}\left[ F_{h}^{\left[ r\right] }\left( \mathbf{r}%
,t\right) \otimes \hat{I}_{h}^{\left[ r\right] }\left( \mathbf{r}\right)
+F_{n}^{\left[ r\right] }\left( \mathbf{r},t\right) \otimes \hat{I}_{n}^{%
\left[ r\right] }\left( \mathbf{r}\right) \right]
\bigr]%
%
\right\}  \label{30}
\end{equation}
In this Eq. (\ref{30}) are present the quantities (generalized fluxes)
\begin{equation}
\hat{I}_{\gamma }^{\left[ r\right] }\left( \mathbf{r\mid }\Gamma \right)
=\int d^{3}p\ u_{\gamma }^{\left[ r\right] }\left( \mathbf{p}\right) \ \hat{n%
}_{1}\left( \mathbf{r},\mathbf{p}\mid \Gamma \right)
=\sum\limits_{j=1}^{N}\int d^{3}p\ u_{\gamma }^{\left[ r\right] }\left(
\mathbf{p}\right) \ \hat{n}_{1j}\left( \mathbf{r},\mathbf{p}\mid \Gamma
_{j}\right)  \label{31}
\end{equation}
with $r=0,1,2,...$ and $\gamma =h$ or $n$, and where
\begin{equation}
\hat{n}_{1j}\left( \mathbf{r},\mathbf{p}\mid \Gamma _{j}\right) =\delta
\left( \mathbf{r}-\mathbf{r}_{j}\right) \delta \left( \mathbf{p}-\mathbf{p}%
_{j}\right)  \label{31a}
\end{equation}
is the individual one-particle dynamical operator and $\Gamma _{j}$
indicates the one-particle phase point $\left( \mathbf{r}_{j},\mathbf{p}%
_{j}\right) $, and
\begin{equation}
u_{n}^{\left[ r\right] }\left( \mathbf{p}\right) =\left[ \mathbf{u}\left(
\mathbf{p}\right) ...\left( r-times\right) ...\mathbf{u}\left( \mathbf{p}%
\right) \right] \qquad ,  \label{32}
\end{equation}
\begin{equation}
u_{h}^{\left[ r\right] }\left( \mathbf{p}\right) =\left( p^{2}/2m\right) \
u_{n}^{\left[ r\right] }\left( \mathbf{p}\right) \qquad .  \label{33}
\end{equation}
Moreover, in Eq. (\ref{32})$\ \left[ \mathbf{u}\left( \mathbf{p}\right)
...\left( r-times\right) ...\mathbf{u}\left( \mathbf{p}\right) \right] $
indicates the tensorial product of $r$-times the generating velocity
\begin{equation}
\mathbf{u}\left( \mathbf{p}\right) =\mathbf{p}/m\qquad ,  \label{34}
\end{equation}
where $r=0$ stands for the densities of energy, $\hat{h}\left( \gamma
=h\right) $, and particles, $\hat{n}\left( \gamma =n\right) $, and their
fluxes of all orders, i.e. the vectorial ones (currents) for $r=1$, and the
higher order (tensorial) ones for $r\geq 2$; moreover, we have designated
the Lagrange multipliers by $\beta ,A,\mathbf{\nu }_{\gamma }\mathbf{,}%
F_{\gamma }^{\left[ r\right] }$, and we recall that $\otimes $ stands for
fully contracted product of tensors. For an alternative derivation of the
auxiliary operator of Eq. (\ref{30}), through the use of the auxiliary one
of Eq. (\ref{29}), see Appendix \textbf{B}.

The average of the quantities of Eq. (\ref{31}) over the nonequilibrium
ensemble provides the macroscopic variables that are the basic ones for the
construction of a nonlinear higher-order hydrodynamics\ (for example, in a
linear approximation, see \cite{JCVMVL02}). The single-particles are
independent and then the statistical operator of Eq. (\ref{29}) can be
written as a product of distributions for each particle, namely
\begin{equation}
\bar{\varrho}\left( \mathbf{\Gamma }\mid t,0\right) =\prod\limits_{j=1}^{N}%
\bar{\varrho}_{j}\left( \mathbf{\Gamma }_{j}\mid t,0\right) \qquad ,
\label{34c}
\end{equation}
with
\begin{equation}
\bar{\varrho}_{j}\left( \mathbf{\Gamma }_{j}\mid t,0\right)
=Z_{j}^{-1}\left( t\right) \exp \left\{ -\int d^{3}r\ d^{3}p\ \varphi
_{1}\left( \mathbf{r},\mathbf{p};t\right) \ \hat{n}_{1j}\left( \mathbf{r},%
\mathbf{p}\mid \Gamma _{j}\right) \right\} \qquad ,  \label{34d}
\end{equation}
where $\hat{n}_{1j}$ is the one-particle dynamical operator of Eq. (\ref{31a}%
), Z$_{j}$ ensures the normalization, $\varphi _{1}$ is the corresponding
Lagrange multiplier. In the grand-canonical description we have that
\[
\bar{\varrho}_{1GC}\left( \mathbf{\Gamma }_{j}\mid t,0\right)
=Z_{j}^{-1}\left( t\right) \exp \left\{ -\int d^{3}r\left[ \beta \left(
\mathbf{r},t\right) \Delta \hat{h}_{1}\left( \mathbf{r}\right) +A\left(
\mathbf{r},t\right) \Delta \hat{n}_{1}\left( \mathbf{r}\right) +\right.
\right.
\]
\[
+\mathbf{\nu }_{h}\left( \mathbf{r},t\right) \cdot \Delta \mathbf{\hat{I}}%
_{h1}\left( \mathbf{r}\right) +\mathbf{\nu }_{n}\left( \mathbf{r},t\right)
\cdot \Delta \mathbf{\hat{I}}_{n1}\left( \mathbf{r}\right) +
\]
\begin{equation}
+\left. \left. \sum\limits_{r\geq 2}\left[ F_{h}^{\left[ r\right] }\left(
\mathbf{r},t\right) \otimes \Delta \hat{I}_{h1}^{\left[ r\right] }\left(
\mathbf{r}\right) +F_{n}^{\left[ r\right] }\left( \mathbf{r},t\right)
\otimes \Delta \hat{I}_{n1}^{\left[ r\right] }\left( \mathbf{r}\right) %
\right] \right] \right\} \qquad ,  \label{34e}
\end{equation}
where
\begin{equation}
\Delta \hat{h}_{1}\left( \mathbf{r}\right) =\hat{h}_{1}\left( \mathbf{r}%
\right) -Tr\left\{ \hat{h}_{1}\left( \mathbf{r}\right) \bar{\varrho}%
_{1}\left( \mathbf{\Gamma }_{j}\mid t,0\right) \right\} \qquad ,  \label{34f}
\end{equation}
\begin{equation}
\hat{h}_{1}\left( \mathbf{r}\right) =\int dp^{3}\frac{p^{2}}{2m}\hat{n}%
_{11}\left( \mathbf{r},\mathbf{p}\mid \Gamma _{j}\right) \qquad ,
\label{34g}
\end{equation}
and similarly for $\Delta \hat{n}_{1}\left( \mathbf{r}\right) $, etc.

But let us now consider -- as it was in the case of the preceding Section
\textbf{3} -- the situation in which we are forced to resort to a truncated
description using, say, a finite small number $s$ of fluxes (that is $%
r=0,1,2,...,s$; $s=1$ in Section \textbf{3}). Therefore Fisher's criterion
of sufficiency is then violated if we proceed with the conventional
statistics. Hence, if we want to analyze an experiment where these $s$
fluxes are accessible, calculating expected values and response function
depending on them, we should introduce USM. We proceed next to analyze such
incomplete description of the hydrodynamics of the ideal gas in USM,
resorting again to Renyi approach \cite{Ren70,JA02}.

Consequently, for obtaining the one-particle statistical operator, $\bar{%
\varrho}_{1}\left( \mathbf{\Gamma }_{j}\mid t,0\right) $ one proceeds to
maximize Renyi generating functional in terms of escort probabilities as
discussed in \textbf{I}, that is [cf. Eq. (\textbf{I}20)]
\begin{equation}
n_{11}\left( \mathbf{r},\mathbf{p};t\right) =\int d\Gamma _{j}\ \hat{n}%
_{1}\left( \mathbf{r},\mathbf{p}\mid \Gamma _{j}\right) \ \stackrel{-}{%
\mathcal{D}_{\alpha }}\left\{ \bar{\varrho}_{1}\left( \mathbf{\Gamma }%
_{j}\mid t,0\right) \right\} \qquad ,  \label{34a}
\end{equation}
which for $\alpha =1$, i.e. when the condition of sufficiency is satisfied,
goes over the conventional expression, as given by Eq. (\textbf{I}.5).

The heterotypical probability distribution is built in terms of the
resulting auxiliary one, the latter, in the grand-canonical description,
given in this case by
\[
\bar{\varrho}_{1\alpha }\left( \Gamma _{j}\mid t,0\right) =\frac{1}{\bar{\eta%
}_{1\alpha }\left( t\right) }\left[ 1+\left( \alpha -1\right) \int d^{3}r\
d^{3}p%
\bigl[%
%
\beta _{\alpha }\left( \mathbf{r},t\right) \ \Delta \hat{h}_{1}\left(
\mathbf{r}\right) +A_{\alpha }\left( \mathbf{r},t\right) \ \Delta \hat{n}%
_{1}\left( \mathbf{r}\right) +\right.
\]
\[
+\mathbf{\nu }_{h\alpha }\left( \mathbf{r},t\right) \cdot \Delta \mathbf{%
\hat{I}}_{h1}\left( \mathbf{r}\right) +\mathbf{\nu }_{n\alpha }\left(
\mathbf{r},t\right) \cdot \Delta \mathbf{\hat{I}}_{n1}\left( \mathbf{r}%
\right)
\]
\begin{equation}
\left. +\sum\limits_{r=2}^{s}\left[ F_{h\alpha }^{\left[ r\right] }\left(
\mathbf{r},t\right) \otimes \Delta \hat{I}_{h1}^{\left[ r\right] }\left(
\mathbf{r}\right) +F_{n\alpha }^{\left[ r\right] }\left( \mathbf{r},t\right)
\otimes \Delta \hat{I}_{n1}^{\left[ r\right] }\left( \mathbf{r}\right) %
\right]
\bigr]%
%
\right\} ^{\frac{1}{1-\alpha }}\qquad ,  \label{34b}
\end{equation}
where $\Delta \hat{h}_{1}$, $\Delta \hat{n}_{1}$, etc., are those defined
after Eq. (\ref{34g}). The distribution of Eq. (\ref{34b}) can be written in
an alternative form using the redefinitions of the Lagrange multipliers
given by
\begin{equation}
\tilde{\beta}_{\alpha }\left( \mathbf{r},t\right) =\beta _{\alpha }\left(
\mathbf{r},t\right) /G_{\alpha }\left( t\right) \qquad ,  \label{35}
\end{equation}
\begin{equation}
\tilde{A}_{\alpha }\left( \mathbf{r},t\right) =A_{\alpha }\left( \mathbf{r}%
,t\right) /G_{\alpha }\left( t\right) \qquad ,  \label{36}
\end{equation}
\begin{equation}
\mathbf{\tilde{\nu}}_{h\alpha }=\mathbf{\nu }_{h\alpha }/G_{\alpha }\left(
t\right) \qquad ,  \label{36a}
\end{equation}
\begin{equation}
\mathbf{\tilde{\nu}}_{n\alpha }=\mathbf{\nu }_{n\alpha }/G_{\alpha }\left(
t\right)  \label{36b}
\end{equation}
and
\begin{equation}
\tilde{F}_{h\alpha }^{\left[ r\right] }\left( \mathbf{r},t\right)
=F_{h\alpha }^{\left[ r\right] }\left( \mathbf{r},t\right) /G_{\alpha
}\left( t\right)  \label{37}
\end{equation}
where
\[
G_{\alpha }\left( t\right) =1-\left( \alpha -1\right) \int d^{3}r\ d^{3}p%
\bigl[\beta _{\alpha }\left( \mathbf{r},t\right) h\left( \mathbf{r},t\right)
+A_{\alpha }\left( \mathbf{r},t\right) n\left( \mathbf{r},t\right) +
\]
\[
+\mathbf{\nu }_{h\alpha }\left( \mathbf{r},t\right) \cdot \mathbf{I}%
_{h}\left( \mathbf{r},t\right) +\mathbf{\nu }_{n\alpha }\left( \mathbf{r}%
,t\right) \cdot \mathbf{I}_{n}\left( \mathbf{r},t\right) +
\]
\begin{equation}
+\sum\limits_{r=2}^{s}\left[ F_{h\alpha }^{\left[ r\right] }\left( \mathbf{r}%
,t\right) \otimes I_{h}^{\left[ r\right] }\left( \mathbf{r},t\right)
+F_{n\alpha }^{\left[ r\right] }\left( \mathbf{r},t\right) \otimes I_{n}^{%
\left[ r\right] }\left( \mathbf{r},t\right) \right] \bigr]\quad ,
\label{37a}
\end{equation}
to obtain that
\begin{equation}
\bar{\varrho}_{1\alpha }\left( \Gamma _{j}\mid t,0\right) =\frac{1}{\bar{\eta%
}_{1\alpha }\left( t\right) }\left[ 1+\left( \alpha -1\right) \int d^{3}r\
d^{3}p\ \xi _{\alpha }\left( \mathbf{r},\mathbf{p};\Gamma _{j}\mid t\right) %
\right] ^{\frac{1}{1-\alpha }}\qquad ,  \label{38}
\end{equation}
where
\begin{equation}
\int d^{3}r\ d^{3}p\ \xi _{\alpha }\left( \mathbf{r},\mathbf{p};\Gamma
_{j}\mid t\right) =\Omega _{j\alpha }\left( \mathbf{r}_{j},\mathbf{p}%
_{j}\mid t\right) \qquad ,  \label{39}
\end{equation}
with
\[
\Omega _{j\alpha }\left( \mathbf{r}_{j},\mathbf{p}_{j}\mid t\right) =\tilde{A%
}_{\alpha }\left( \mathbf{r}_{j},t\right) +\tilde{\beta}_{\alpha }\left(
\mathbf{r}_{j},t\right) \frac{p_{j}^{2}}{2m}+\mathbf{\tilde{\nu}}_{n\alpha
}\left( \mathbf{r}_{j},t\right) \cdot \frac{\mathbf{p}_{j}}{m}+
\]
\begin{equation}
+\mathbf{\tilde{\nu}}_{h\alpha }\left( \mathbf{r}_{j},t\right) \cdot \frac{%
p_{j}^{2}}{2m}\frac{\mathbf{p}_{j}}{m}+\sum\limits_{r=2}^{s}\sum\limits_{%
\gamma =n\ or\ h}\tilde{F}_{\gamma \alpha }\left( \mathbf{r}_{j},t\right)
\otimes u_{\gamma }^{\left[ r\right] }\left( \mathbf{p}_{j}\right)
\label{40}
\end{equation}
after using the definitions of the densities and fluxes of order $r=1$ to $s$
in terms of the individual single-particle dynamical function of Eq. (\ref
{27}), and the integrations in \textbf{r} and \textbf{p} are performed..

Consequently,\ the auxiliary probability density of Eq. (\ref{34c}) becomes
the product of factors involving each particle individually, given in the
Renyi%
\'{}%
s approach we used by
\begin{equation}
\bar{\varrho}_{\alpha }\left( \Gamma \mid t,0\right) \simeq \frac{1}{\bar{%
\eta}_{\alpha }\left( t\right) }\prod\limits_{j=1}^{N}\left[ 1+\left( \alpha
-1\right) \Omega _{j\alpha }\left( \Gamma \mid t\right) \right] ^{\frac{1}{%
1-\alpha }}\qquad .  \label{46}
\end{equation}

In section \textbf{3} we have applied these results, for $s=1$, to derive\
an ``anomalous'' diffusion equation (for fitting the experimental data) in
voltammetry measurements in microbatteries with fractal electrodes, and in
continuation we present an additional application consisting in a study of
hydrodynamic properties of an ideal gas of photons in black-body radiation,
extending these results to deal now with a quantum system.

\section{RADIATION IN INSUFFICIENT DESCRIPTION}

Let us consider an ideal gas of photons in the presence of an uniform flux
of energy (generated, for example, having different temperatures at both
ends of the container). We look for the determination of the energy, what is
done, on the one side using the conventional approach in a description that
includes the energy and the energy flux as basic variables and, on the other
side, using a quite incomplete description which includes only energy, but
dealt with in USM. We indicated by $\omega _{\mathbf{k}}=c\ \mathbf{k}$ the
photon frequency-dispersion relation, and by $a_{\mathbf{k}}\left( a_{%
\mathbf{k}}^{\dagger }\right) $ the annihilation (creation) operators in
states $\left| \mathbf{k}\right\rangle $. The energy and flux of energy
dynamical operators are given by
\begin{equation}
\hat{H}=\sum\limits_{\mathbf{k}}\hslash \omega _{\mathbf{k}}\ a_{\mathbf{k}%
}^{\dagger }a_{\mathbf{k}}\qquad ,  \label{47}
\end{equation}
\begin{equation}
\mathbf{\hat{I}}_{h}=\sum\limits_{\mathbf{k}}\hslash \omega _{\mathbf{k}}\
\nabla _{\mathbf{k}}\omega _{\mathbf{k}}\ a_{\mathbf{k}}^{\dagger }a_{%
\mathbf{k}}\qquad .  \label{48}
\end{equation}

The auxiliary (``instantaneous frozen quasiequilibrium'') statistical
operator including both dynamical variables in the conventional formalism is
\begin{equation}
\bar{\varrho}\left( t,0\right) =Z^{-1}\left( t\right) \exp \left\{
-F_{h}\left( t\right) \hat{H}-\mathbf{\nu }_{h}\left( t\right) \cdot \mathbf{%
\hat{I}}_{h}\right\} \qquad ,  \label{49}
\end{equation}
where $F_{h}$ and $\mathbf{F}_{h}$ are the associated Lagrange multipliers
(intensive nonequilibrium thermodynamical variable), with the nonequilibrium
statistical operator given, in terms of this $\bar{\varrho}$, by Eq. (%
\textbf{A}.14) to (\textbf{A}.17). The averages of the basic dynamical
variables can be calculated to obtain respectively that
\begin{equation}
E\left( t\right) =\sum\limits_{\mathbf{k}}\hslash \omega _{\mathbf{k}}\ \nu
_{\mathbf{k}}\left( t\right) \qquad ,  \label{50}
\end{equation}
\begin{equation}
\mathbf{I}_{h}\left( t\right) =\sum\limits_{\mathbf{k}}\hslash \omega _{%
\mathbf{k}}\ \nabla _{\mathbf{k}}\omega _{\mathbf{k}}\ \nu _{\mathbf{k}%
}\left( t\right) \qquad ,  \label{51}
\end{equation}
where
\begin{equation}
\nu _{\mathbf{k}}\left( t\right) =\left[ \exp \left\{ F_{h}\left( t\right)
\hslash \omega _{\mathbf{k}}+\mathbf{\nu }_{h}\left( t\right) \cdot \hslash
\omega _{\mathbf{k}}\ \nabla _{\mathbf{k}}\omega _{\mathbf{k}}\right\} -1%
\right] ^{-1}  \label{52}
\end{equation}

Considering the presence of a weak flux, introducing an expansion in $%
\mathbf{F}_{h}\left( t\right) $ keeping only the first nonnull contribution,
we obtain the equations of state (i.e. relation between the basic
nonequilibrium thermodynamic variables and the intensive one) given by
\begin{equation}
\frac{1}{V}\ E\left( t\right) =a\ F_{h}^{-4}\left( t\right) +a_{s}\left(
t\right) \ \left| \mathbf{\nu }_{h}\left( t\right) \right| ^{2}\qquad ,
\label{53}
\end{equation}
\begin{equation}
\frac{1}{V}\ \mathbf{I}_{h}\left( t\right) =b\left( t\right) \ \mathbf{\nu }%
_{h}\left( t\right) \qquad ,  \label{54}
\end{equation}
where
\begin{equation}
a=\pi ^{2}/15\hslash ^{3}c^{3}\qquad ,  \label{55}
\end{equation}
\begin{equation}
a_{s}\left( t\right) =2\pi ^{2}/9\hslash ^{3}cF_{h}^{6}\left( t\right)
\qquad ,  \label{56}
\end{equation}
\begin{equation}
b\left( t\right) =-4\pi ^{2}/45\hslash ^{3}cF_{h}^{5}\left( t\right)
\label{57}
\end{equation}
as shown in Appendix \textbf{C}. It has been used the relation $\omega _{%
\mathbf{k}}=c\ \mathbf{k}$, and we notice that in equilibrium $\mathbf{\nu }%
_{h}=0$ and $F_{h}=\beta =\left[ k_{B}T\right] ^{-1}$, and then Eq. (\ref{53}%
) becomes Stefan-Boltzmann law.

Using Eq. (\ref{54}) in Eq. (\ref{53}) we can write for the energy that
\begin{equation}
\frac{1}{V}\ E\left( t\right) =\frac{a}{F_{h}^{4}\left( t\right) }+\frac{15}{%
8ac^{2}}F_{h}^{4}\left( t\right) \left| \frac{\mathbf{I}_{h}\left( t\right)
}{V}\right| ^{2}\qquad .  \label{58}
\end{equation}

Let us next consider the most insufficient description, done in terms of
only the energy, but using USM in Renyi's approach, when the auxiliary
statistical operator is given by
\begin{equation}
\bar{\varrho}_{\alpha }\left( t,0\right) =\frac{1}{\bar{\eta}_{\alpha
}\left( t\right) }\left[ 1+\left( \alpha -1\right) \varphi _{h}\left(
t\right) \ \Delta \hat{H}\right] ^{\frac{1}{\alpha -1}}\qquad ,  \label{59}
\end{equation}
with $\Delta \hat{H}=\hat{H}-E\left( t\right) $, and we recall that in the
calculation of average values is used the escort probability expressed in
terms of $\bar{\varrho}_{\alpha }$ [cf. Eq. (\ref{15a})]. On the given
assumption that the flux is weak, one should expect that $\alpha $ is close
to $1$, we write $\alpha =1+\epsilon $ (or $\epsilon =\alpha -1$) and we
perform the calculations retaining only terms up to first order in $\epsilon
$. As shown in Appendix \textbf{C}, after some calculation we find that
\begin{equation}
E\left( t\right) =E_{o}\left( t\right) -\left( \alpha -1\right) \varphi
_{h}\left( t\right) \ \sigma _{E}^{2}\left( t\right) \qquad ,  \label{60}
\end{equation}
where
\[
E_{o}\left( t\right) =\sum\limits_{\mathbf{k}}\hslash \omega _{\mathbf{k}}\
\nu _{\mathbf{k}}\left( t\right) =
\]
\begin{equation}
=V\ \Gamma \left( 4\right) \ \zeta \left( 4\right) /2\pi ^{2}\ \left(
\hslash c\right) ^{3}\varphi _{h}^{4}\left( t\right) =\frac{a}{%
F_{h}^{4}\left( t\right) }\qquad ,  \label{61}
\end{equation}
with $a$ given in Eq. (\ref{55}), $\Gamma $ and $\zeta $ being Gamma and
Riemann functions respectively, which follow after using that
\begin{equation}
\mathcal{N}_{\mathbf{k}}\left( t\right) =\left[ \exp \left\{ \varphi
_{h}\left( t\right) \ \hslash \omega _{\mathbf{k}}\right\} -1\right]
^{-1}\qquad ,  \label{62}
\end{equation}
and $\sigma _{E}^{2}\left( t\right) $ is the fluctuation of energy
\[
\sigma _{E}^{2}\left( t\right) =\sum\limits_{\mathbf{k}}\left( \hslash
\omega _{\mathbf{k}}\right) ^{2}\ \mathcal{N}_{\mathbf{k}}\left( t\right) \ %
\left[ 1+\mathcal{N}_{\mathbf{k}}\left( t\right) \right] -\left[
\sum\limits_{\mathbf{k}}\hslash \omega _{\mathbf{k}}\ \mathcal{N}_{\mathbf{k}%
}\left( t\right) \right] ^{2}=
\]
\begin{equation}
=4\left( \alpha -1\right) \left[ V\ \Gamma \left( 4\right) \ \zeta \left(
4\right) /2\pi ^{2}\ \left( \hslash c\right) ^{3}\varphi _{h}^{4}\left(
t\right) \right] \qquad ,  \label{63}
\end{equation}
and then
\begin{equation}
\frac{1}{V}\ E\left( t\right) =a\ \varphi _{h}^{-4}\ \left[ 1-4\left( \alpha
-1\right) \right] \qquad .  \label{64}
\end{equation}
Equating Eqs. (\ref{58}) and (\ref{64}) there follows an equation for the
infoentropic index $\alpha $ in terms of the variables which characterize
the macroscopic state of the system, namely
\begin{equation}
\frac{a}{\varphi _{h}^{4}}\ \left[ 1-4\left( \alpha -1\right) \right] =\frac{%
a}{\varphi _{h}^{4}}\ +\frac{15}{8}\ \frac{\varphi _{h}^{4}}{ac^{2}}\left[
\frac{I}{V}\right] ^{2}\qquad .  \label{65}
\end{equation}

But in the noted condition of weak flux we admit that $F_{h}\simeq \beta
_{o} $ and $\varphi _{h}\simeq \beta _{o}$ with, we recall, $\epsilon
=\alpha -1$, and where $\beta _{o}$ is the reciprocal of the average
temperature of the gas (its gradient being small). Hence, using these
results in Eq. (\ref{65}) we arrive at the expression for $\alpha $ given
approximately by
\begin{equation}
\alpha \simeq 1-\mathcal{C}\left[ \frac{I}{V}\right] ^{2}\qquad ,  \label{66}
\end{equation}
where
\begin{equation}
\mathcal{C}=\frac{15}{8}\frac{\beta _{o}^{8}}{a^{2}c^{2}}\qquad .  \label{67}
\end{equation}

It can be noticed that $\mathcal{C}^{-1}\sim \left( a\beta _{o}^{-4}c\right)
^{2}$ is something like the square of a flux of energy composed of the
energy density of the radiation, $a\beta _{o}^{-4}$, traversing at the speed
of light, while we should expect $I$ to be composed of something like the
density of energy traversing at a speed determined by the gradient of
temperature, and then, in fact, we do have that for the ideal gas of photons
(black-body radiation) works quite well meaning that the insufficiency of
characterization can be ignored.

\section{IDEAL GAS IN A FINITE BOX}

We consider in this section other observation concerning the ideal gas,
namely, that as a rule work in the thermodynamic limit is usually taken,
what implies an infinite volume, but real systems are finite \cite{Lagos}.
Then in the case of an ideal quantum gas in a finite box of volume $V$ and
area $A$, summation over the states $\mathbf{k}\equiv \left[ n_{x}\left( \pi
/L_{x}\right) ;n_{y}\left( \pi /L_{y}\right) ;n_{z}\left( \pi /L_{z}\right) %
\right] $ (where $n_{x},n_{y},n_{z}$ are $1,2$, etc. and $L_{x},L_{y},L_{z}$
are the size of the sides of the box) can be exactly replaced by an integral
only in the thermodynamic limit. Otherwise we do have corrections depending
on the area of the surface, and in the calculation of the partition
function, internal energy, etc., there appears nonextensive terms.

According to Pathria's textbook \cite{Pat96}, and see also \cite{Pat98}, the
density of states of the ideal gas in the finite box is given by
\begin{equation}
g\left( \epsilon \right) =V\left[ 2\pi \left( m/2\pi ^{2}\hslash ^{2}\right)
^{3/2}\epsilon ^{1/2}-\frac{1}{8}\frac{A}{V}\left( m/\pi \hslash ^{2}\right)
+...\right] \qquad ,  \label{68}
\end{equation}
where $V$ is the volume and $A$ the area, and dots stand for contributions
quadratic and of higher powers in $\left( A/V\right) $; once we consider a
finite but large box it is kept in what follows only the first-order
contribution in $\left( A/V\right) $.

Taking the equilibrium grand-canonical ensemble with temperature $T$ and
chemical potential $\mu $, using the density of states of Eq. (\ref{68})\
and taking for simplicity the statistically nondegenerate condition, it
follows that (see Appendix \textbf{D})
\begin{equation}
\frac{E}{N}\simeq \frac{3}{2}k_{B}T\ \left[ 1-\frac{1}{6}\frac{A\lambda _{T}%
}{V}\right] \lambda _{T}^{-3}\exp \left\{ \mu /k_{B}T\ \right\} \qquad ,
\label{69}
\end{equation}
for the density of energy, and
\begin{equation}
\frac{N}{V}\simeq \left[ 1-\frac{1}{4}\frac{A\lambda _{T}}{V}\right] \lambda
_{T}^{-3}\exp \left\{ \mu /k_{B}T\ \right\} \qquad ,  \label{70}
\end{equation}
for the density of particles; $\lambda _{T}$ is the mean thermal de Broglie
wavelength, namely, $\lambda _{T}^{2}=\hslash ^{2}/mk_{B}T$. Hence, the
energy per particle is given by
\begin{equation}
\frac{E}{N}\simeq \frac{3}{2}k_{B}T\ \left[ 1+\frac{1}{12}\frac{A\lambda _{T}%
}{V}\right] \qquad .  \label{71}
\end{equation}
Evidently, in the thermodynamic limit (infinite box) it follows the standard
result $E=\left( 3/2\right) Nk_{B}T$ (we recall that we took the
nondegenerate condition).

Next we admit to be in the condition of insufficiency consisting in ignoring
the finite size of the box, and we proceed with the calculations in the
thermodynamics limit but resorting -- to patch the limitation thus
introduced -- to Unconventional Statistical Mechanics.

Using USM in Renyi's approach, in the insufficient condition, namely, using
the thermodynamic limit (infinite-size box), that is using in the
calculations the escort probability [cf. Eq. (\ref{15a})] in terms of the
auxiliary Renyi's heterotypical distribution
\begin{equation}
\bar{\varrho}_{\alpha }=\frac{1}{\bar{\eta}_{\alpha }}\left\{ 1+\left(
\alpha -1\right) \left[ F_{h}\left( \hat{H}-\left\langle \hat{H}%
\right\rangle \right) +F_{n}\left( \hat{N}-\left\langle \hat{N}\right\rangle
\right) \right] \right\} ^{\frac{1}{1-\alpha }}\qquad ,  \label{72}
\end{equation}
and under the expected condition of $\alpha $ near $1$ (i.e. $\left| \alpha
-1\right| \ll 1$ for a very large box it follows that
\begin{equation}
E\simeq E_{o}-\left( \alpha -1\right) \left[ F_{h}\ \sigma _{E}^{2}+F_{n}\
\sigma _{EN}^{2}\right] \qquad ,  \label{73}
\end{equation}
\begin{equation}
N=N_{o}-\left( \alpha -1\right) \left[ F_{n}\ \sigma _{N}^{2}+F_{h}\ \sigma
_{NE}^{2}\right] \qquad ,  \label{74}
\end{equation}
where $\sigma _{E}^{2}$ and $\sigma _{N}^{2}$ are the energy and
particle-number correlation functions, and $\sigma _{EN}^{2}$ and $\sigma
_{NE}^{2}$ the cross-correlation functions (see Appendix \textbf{D}, for
details) and where $E_{o}=\left( 3N_{o}/2F_{h}\right) $ and $N_{o}$ are the
first contribution in the expansion around $\alpha =1$.

Finally, we can write (see Appendix \textbf{D)}
\begin{equation}
\frac{E}{N}\simeq \frac{3}{2}F_{h}^{-1}\ \left[ 1+\frac{1}{4}\left( \alpha
-1\right) \right] \qquad ,  \label{75}
\end{equation}
where, evidently, in the thermodynamic limit $\alpha $ approaches $1$ as it
should. Moreover, from Eq. (\ref{75}) we can clearly see that the
infoentropic index $\alpha $ depends on the system dynamics, its geometry
and size, and the thermodynamic state.

Further applications of USM are available in the literature on the subject,
and we can mention the interesting cases of its use on dealing with
hydrodynamic turbulence and collider physics in \cite{Bec02} and \cite{Bec00}
respectively.

\section{FINAL REMARKS}

We have illustrated in this paper the use of what we have called
Unconventional Statistical Mechanics, an auxiliary formalism which can
provide a theoretical approach to situations when the conventional, and well
established Boltzmann-Gibbs statistical mechanics has its use impaired
because of a lack of a proper knowledge (for the problem in hands) of the
characterization of the system and its dynamics on the part of the
researcher (the Fisher's criterion of sufficiency is not satisfied as
discussed in \cite{LVR02}).

We have here dealt with \textit{``anomalous'' luminescence} from quantum
wells in semiconductor heterostructures, where the \textit{failure of
sufficiency} resides at a microscopic level, in that one does not know the
proper quantum mechanical states in the thin (nanometric scale) quantum
well. This is the result that we cannot precisely solve Schr\"{o}dinger
equation for the carriers because a failure to impose boundary conditions on
the fractal-like morphological structure of the boundaries, to which we do
not have easy access; evidently the wave function is largely affected by the
roughness of the frontiers in the nanometric structure. As shown, as the
width of the quantum well increases, the ``anomaly'' tends to disappear.
This is an example in the realm of semiconductor physics.

A second illustration, the one in section \textbf{3}, is in the area of
electro-physico-chemistry, namely the behavior of thin electrodes with a
fractal-like morphology in microbatteries. An analysis of experiments of the
so-called cyclic voltammetry has been performed, whose results are a
consequence of the occurrence of an \textit{``anomalous'' diffusion} of
charges from the electrolyte. In this case the \textit{failure of sufficiency%
} resides at a macroscopic, or, better to say, mesoscopic level, when
classical Onsagerian hydrodynamics is used instead of the higher-order
hydrodynamics that the problem requires. As described the ``anomaly''
depends on the experimental protocol, and tends to disappear as the rate of
charge transfer is increased.

The third illustration, see section \textbf{4}, is a purely theoretical
analysis concerning ``anomalous'' statistics of an ideal gas. As shown, once
the ideal gas is perfectly and completely characterized in terms of the
one-particle density function (Dirac-Landau-Wigner single-particle dynamical
operator in the quantum case), the only possible description is the
conventional one. But, if instead we work with truncated sets of linear
combinations of the one-particle density, viz. the densities of particles
and energy and the set of their fluxes up to a certain order, in a partial
generalized grand-canonical description (and then we do have insufficiency
of description), \ to overcome the fact that we are not complying with the
criterion of sufficiency (i.e. to calculate properties) unconventional
statistics can be introduced, at the price of having undetermined
parameter(s). In terms of them one obtains ``anomalous'' hydrodynamic laws,
for example, in the lowest order approximation, ``anomalous'' diffusion,
with a power law, in the subsequent order an ``anomalous'' Maxwell-Cattaneo
equation (damped waves), and so on. As the order used in the higher-order
hydrodynamics increases the ``anomaly'' tends to disappear because one is
approaching a description equivalent to the one in terms of the
single-particle density matrix.

It can be noticed that we have always written the word anomalous within
quotation marks: This is so because there is nothing anomalous in the
physical laws governing the system: what is ``anomalous'' is the result
obtained using the unconventional approach when, as noticed, we are unable
to comply with the \textit{criterion of sufficiency} in the use of the
proper Boltzmann-Gibbs statistics. Moreover, concerning the, what we have
called, \textit{``path to sufficiency''} evidenced in the three cases we
presented, it is tempting to conjecture that it may be a general
characteristic of unconventional statistical mechanics, in any of the cases
when different infoentropic-index-dependent structured informational\
entropies are used.

Finally, we add a couple of comments concerning other possible situations
that can be analyzed in terms of USM, and its connection with the criteria
of efficiency and sufficiency.

Application of a USM resorting to Havrda-Charvat structural infoentropy to
the question of the evolution of a nonequilibrium temperature-like variable,
say $T^{\ast }\left( t\right) $, in the case of a system with long-range
spatial correlations \cite{LR02} shows that in the approach to equilibrium
is present a long and near-stationary plateaux above the reservoir
temperature which is to be attained in the long range once final thermal
equilibrium follows. The calculations are not related (compared) to
experimental results, and so we do not know if they have any meaning, except
to leave a suggestion of theoretical results in search of an experiment. If
the experiment shall show in fact that there are differences in relation to
the conventional calculation (``anomalous'' results) one should look where
Fisher's criteria of efficiency and sufficiency is not satisfied (if it is
the case, as already noticed should be related to improper handling of the
long-range correlations; see below the case of the one-electron transistor).
There exists a situation in the physics of semiconductors where is also
present a near stationary state in the way to equilibrium associated to a
slowing-down of the relaxation processes. In the case of the out of
equilibrium photoinjected double plasma in semiconductors (see for example
Chapter \textbf{6} in \cite{LVR1}) what is called the carriers'
quasitemperature $T^{\ast }\left( t\right) $ follows a path to equilibrium
which strongly depends on the experimental protocol \cite{PAVL01}. Thermal
equilibrium with the reservoir follows very rapidly (tens to a hundred
picoseconds) when the exciting laser pulse is short (pico- to subpico-second
scale), but is slow and presenting a long near plateaux when the exciting
lase pulse is long (tens of picoseconds duration). Such behavior in the
latter case is not properly described by theory in the conventional
formalism but could, we think, as in \cite{LR02} be obtained resorting to
USM with an appropriate adjustment of the infoentropic index involved in,
for example, Havrda-Charvat or Renyi approaches. The original failure in
this case consisted that in the treatment that was applied one does not
comply with the \textit{criterion of sufficiency} by not considering the
facts responsible for the phenomenon to occur, namely, ``phonon bottleneck''
and ambipolar diffusion (the latter keeps decreasing the density of
particles), which once incorporated in the description produces an excellent
agreement of (conventional) theory and experiment, and we have sufficiency
at work \cite{PAVL01}.

Another interesting case to be considered is the one of single-electron
transistors \cite{LNO}, when long-range and strong (because not screened)
Coulomb interaction leads to difficulties when dealt with in a simple way in
the conventional approach. Again, the point is a failure of sufficiency: in
this case is the proper characterization of the states of the system as a
result of long-range interconnected correlations. When renormalization of
the carrier states is introduced, sufficiency is restored and there follows
an excellent agreement between (conventional) theory and experimental data
\cite{LNO}. We conjecture that instead of application of the renormalization
group for, as said, restore sufficiency (i.e. a good mimical description of
the carriers' states), one could use the unrenormalized description in terms
of USM (a point to be considered in the future).

In the last two cases above it is also conjecturable that there may be
present a kind of ``path to sufficiency'', of the like of those presented in
the three other former examples.

In a concluding remark, we can say, in summary, that the illustrations here
presented of the application of the theory of the preceding article, allows
for gaining a better perspective of USM, which appears as a \textit{useful
and practical formalism for the macroscopic description of systems when the
research cannot comply with criteria of efficiency and/or sufficiency in the
conventional, well established, and physically and logically sound
Boltzmann-Gibbs theory}. Basically it is a sophisticated fitting formalism
to be used in the referred conditions, namely, when one does not have a
proper access to the characteristics of the system that are \textit{relevant}
to determine the property of the system we are studying.

It is worth noticing that in the illustrations we presented there appears a
quite interesting connection between theory, experiment, and front-line
technology.\bigskip

{\large \textbf{ACKNOWLEDGMENTS}}

We acknowledge financial support provided to our Group in different
opportunities by the S\~{a}o Paulo State Research Foundation (FAPESP), the
Brazilian National Research Council (CNPq), the Ministry of Planning
(Finep), the Ministry of Education (CAPES), Unicamp Foundation (FAEP), IBM
Brasil, and the John Simon Guggenheim Memorial Foundation (New York, USA).%

\appendix%
%

\renewcommand{\thesection}{Appendix A}%
%

\section{``Anomalous'' Diffusion}

\setcounter{equation}{0}%
%
\renewcommand{\theequation}{A.\arabic{equation}}%
%

Let us consider first the conventional case of diffusion, when the criterion
of sufficiency is satisfied, meaning the several stringent restrictions its
validity requires are met, namely, when are satisfied the conditions of
local equilibrium, linear Onsager relations and symmetry laws, condition
such that the motion is dominated by long wavelengths and very low
frequencies contributions, and weak fluctuations. A specific criterion for
validity is given in Ref. \cite{RVL00b}, where it is considered a
Brownian-like system composed of two ideal fluids in interaction between
them. The continuity equation for the flux $\mathbf{I}_{n}$ (Eq. (17) in
\cite{RVL00b}) after transforming Fourier in time, takes the form
\begin{equation}
\left( 1+i\omega \tau _{n1}\right) \ \mathbf{I}_{n}\left( \mathbf{r},\omega
\right) +\tau _{n1}\nabla \cdot \ I_{n}^{\left[ 2\right] }\left( \mathbf{r}%
,\omega \right) =0\qquad ,  \label{I1}
\end{equation}
that is Eq. (\ref{11}) for $r=1$, and using Eq. (\ref{16}) for the collision
integral and then $\tau _{n1}$ is the momentum relaxation time. But, at
sufficiently low frequencies, meaning that $\omega \tau _{n1}$ can be
neglected, we have the appropriate expression
\begin{equation}
\mathbf{I}_{n}\left( \mathbf{r},\omega \right) \simeq -\tau _{n1}\nabla
\cdot \ I_{n}^{\left[ 2\right] }\left( \mathbf{r},\omega \right) \qquad ,
\label{I2}
\end{equation}
and a direct calculation tells us that
\begin{equation}
\nabla \cdot \ I_{n}^{\left[ 2\right] }\left( \mathbf{r},\omega \right)
\simeq \left( k_{B}T/m\right) \ \nabla n\left( \mathbf{r},\omega \right) \
\qquad .  \label{I3}
\end{equation}

Replacing Eq. (\ref{I3}) in Eq. (\ref{I2}), and the latter in the
conservation equation for the density (Eq. (16) in \cite{RVL00b}), we obtain
the usual Fick's diffusion equation
\begin{equation}
\frac{\partial }{\partial t}n\left( \mathbf{r},t\right) +D\ \nabla
^{2}n\left( \mathbf{r},t\right) =0\qquad ,  \label{I4}
\end{equation}
where $D=\frac{1}{3}v_{th}^{2}\ \tau _{n1}$, with $\frac{1}{2}m\ v_{th}^{2}=%
\frac{3}{2}k_{B}T$; $v_{th}$ is the thermal velocity and $D$ the diffusion
coefficient, with dimension $cm^{2}/\sec $.

Let us now go over the unconventional treatment, which is required once one
is looking forward for an adjustment of data on the basis of a description
in terms of a diffusive movement, when this is not possible, as a
consequence that diffusion in the microroughneessed region is governed by
not too long wavelengths (up to the nanometric ones, i.e. $10^{-7}\ cm$,
while the limitation of the diffusive domain \cite{RVL00b} is the order of $%
D/v_{th}$, say, typically $10^{-2}$ to $10^{-4}\ cm$). A higher-order
hydrodynamics \cite{JCVMVL02} needs be introduced, but if the lower order
description including only the density and its flux is kept, then we are not
complying with the criterion of sufficiency and we need to introduce
Unconventional Statistical Mechanics. Let us consider the auxiliary
statistical operator which for this system of free particles can be
approximated by a product of the statistical operator of the individual
particles, namely (cf. Section \textbf{4})
\begin{equation}
\bar{\varrho}\left( \Gamma \mid t,0\right) =\prod\limits_{j=1}^{N}\bar{%
\varrho}_{1\alpha }\left( \Gamma _{1}\mid t,0\right) \qquad ,  \label{I4a}
\end{equation}
where $\Gamma _{1}$ stands for the one-particle phase point $\left( \mathbf{r%
}_{j},\mathbf{p}_{j}\right) $ and in Renyi statistics we do have that
\begin{equation}
\bar{\varrho}_{1\alpha }\left( \Gamma _{1}\mid t,0\right) =\frac{1}{%
z_{\alpha }\left( t\right) }\left[ 1+\left( \alpha -1\right) \int d^{3}r\int
d^{3}p\ \varphi _{1\alpha }\left( \mathbf{r},\mathbf{p},t\right) \ \Delta
\hat{n}_{11}\left( \mathbf{r},\mathbf{p},t\right) \right] ^{-\frac{1}{\alpha
-1}}\qquad ,  \label{I5}
\end{equation}
where
\begin{equation}
\Delta \hat{n}_{11}\left( \mathbf{r},\mathbf{p},t\right) =\hat{n}_{11}\left(
\mathbf{r},\mathbf{p}\mid \Gamma _{1}\right) -\left\langle \hat{n}%
_{11}\left( \mathbf{r},\mathbf{p}\mid \Gamma _{1}\right) \right\rangle
_{\alpha }\qquad ,  \label{I6}
\end{equation}
with
\[
\hat{n}_{11}\left( \mathbf{r},\mathbf{p}\mid \Gamma _{1}\right) =\delta
\left( \mathbf{r}-\mathbf{r}_{j}\right) \delta \left( \mathbf{p}-\mathbf{p}%
_{j}\right) \qquad ,
\]
\begin{equation}
\left\langle \hat{n}_{11}\left( \mathbf{r},\mathbf{p}\mid \Gamma _{1}\right)
\right\rangle _{\alpha }=\int d\Gamma \ \hat{n}_{11}\left( \mathbf{r},%
\mathbf{p}\mid \Gamma _{1}\right) \ \stackrel{-}{\mathcal{D}}_{1\alpha
}\left\{ \bar{\varrho}_{1\alpha }\left( \Gamma _{1}\mid t,0\right) \right\}
\ \ ,  \label{I7}
\end{equation}
and
\begin{equation}
\varphi _{1\alpha }\left( \mathbf{r},\mathbf{p},t\right) =F_{n\alpha }\left(
\mathbf{r},t\right) +F_{h\alpha }\left( \mathbf{r},t\right) \frac{p^{2}}{2m}+%
\mathbf{F}_{n\alpha }\left( \mathbf{r},t\right) \cdot \frac{\mathbf{p}}{m}
\label{I8}
\end{equation}
are modified forms of the associated Lagrange multiplier which appear in Eq.
(\textbf{I}.16) in \textbf{I}, where in the latter the change in space of
the energy density has been desconsidered (i.e. we took $F_{h\alpha }\left(
\mathbf{r},t\right) =\beta $), $z$ ensures its normalization, and
\begin{equation}
\ \stackrel{-}{\mathcal{D}}_{1\alpha }\left\{ \bar{\varrho}_{1\alpha }\left(
\Gamma _{1}\mid t,0\right) \right\} =\left[ \bar{\varrho}_{1\alpha }\left(
\Gamma _{1}\mid t,0\right) \right] ^{\alpha }/\int d\Gamma _{1}\ \left[ \bar{%
\varrho}_{1\alpha }\left( \Gamma _{1}\mid t,0\right) \right] ^{\alpha
}\qquad ,  \label{I8a}
\end{equation}
is the accompanying \textit{escort probability} (see \textbf{I}; \cite{Ren70}%
)$.$

Introducing the modified Lagrange multiplier
\begin{equation}
\tilde{\varphi}_{1\alpha }\left( \mathbf{r},\mathbf{p},t\right) =\varphi
_{1\alpha }\left( \mathbf{r},\mathbf{p},t\right) /\left[ 1-\left( \alpha
-1\right) \int d^{3}r\int d^{3}p\varphi _{1\alpha }\left( \mathbf{r},\mathbf{%
p},t\right) \ \left\langle \hat{n}_{11}\left( \mathbf{r},\mathbf{p}\mid
\Gamma _{1}\right) \right\rangle _{\alpha }\right] ,  \label{I9}
\end{equation}
we find that
\begin{equation}
\bar{\varrho}_{1\alpha }\left( \Gamma \mid t,0\right) =\frac{1}{\bar{Z}%
\left( t\right) }\left[ 1+\left( \alpha -1\right) \int d^{3}r\int d^{3}p\
\tilde{\varphi}_{1\alpha }\left( \mathbf{r},\mathbf{p},t\right) \ \hat{n}%
_{11}\left( \mathbf{r},\mathbf{p}\mid \Gamma _{1}\right) \right] ^{-\frac{1}{%
\alpha -1}}  \label{I10}
\end{equation}
with $\bar{Z}\left( t\right) $ ensuring the normalization condition.

Using Eq. (\ref{I4a}) to Eq. (\ref{I10}), after some lengthy but
straightforward calculations, it follows for the energy density that
\begin{equation}
h\left( \mathbf{r},t\right) =u\left( \mathbf{r},t\right) +n\left( \mathbf{r}%
,t\right) \frac{1}{2}mv_{\alpha }^{2}\left( \mathbf{r},t\right) \qquad ,
\label{I19}
\end{equation}
i.e., composed of the energy associated to the drift movement (the last
term) and the internal energy density
\begin{equation}
u\left( \mathbf{r},t\right) =\frac{3}{5-3\alpha }\frac{\mathcal{C}_{\alpha
}\left( \mathbf{r},t\right) }{\beta _{\alpha }\left( \mathbf{r},t\right) }%
n^{\gamma _{\alpha }}\left( \mathbf{r},t\right) \qquad ,  \label{I20}
\end{equation}
where
\begin{equation}
\mathcal{C}_{\alpha }\left( \mathbf{r},t\right) =\left\{ \frac{2\pi N}{%
\left( \alpha -1\right) ^{\frac{3}{2}}\bar{Z}\left( t\right) }\left[ \frac{2m%
}{\tilde{\beta}_{\alpha }\left( \mathbf{r},t\right) }\mathcal{B}\left( \frac{%
3}{2},\frac{\alpha }{\alpha -1}-\frac{3}{2}\right) \right] \right\} ^{\frac{%
2\left( \alpha -1\right) }{\alpha -3}}  \label{I21}
\end{equation}
$\mathcal{B}\left( \nu ,x\right) $ is the Beta function, we have written $%
F_{h\alpha }\left( \mathbf{r},t\right) =\beta _{\alpha }\left( \mathbf{r}%
,t\right) $; $F_{n\alpha }\left( \mathbf{r},t\right) =m\beta _{\alpha
}\left( \mathbf{r},t\right) \mathbf{v}_{\alpha }\left( \mathbf{r},t\right) $
introducing a ``drift velocity'' field $\mathbf{v}_{\alpha }\left( \mathbf{r}%
,t\right) $; moreover
\begin{equation}
\gamma _{\alpha }=\frac{3\alpha -5}{\alpha -3}\qquad ,  \label{I23}
\end{equation}
and the values of $\alpha $ are restricted to the interval
\begin{equation}
1\leq \alpha <\frac{5}{3}\qquad .  \label{I23a}
\end{equation}

Finally, the second order flux is given by
\[
I_{n}^{\left[ 2\right] }\left( \mathbf{r},t\right) =\int d^{3}p\left[ \frac{%
\mathbf{p}}{m}\frac{\mathbf{p}}{m}\right] \ \left\langle \hat{n}_{11}\left(
\mathbf{r},\mathbf{p}\mid \Gamma _{1}\right) \right\rangle _{\alpha }=
\]
\begin{equation}
=n\left( \mathbf{r},t\right) \left[ \mathbf{v}\left( \mathbf{r},t\right)
\mathbf{v}\left( \mathbf{r},t\right) \right] +\frac{2}{3m}u\left( \mathbf{r}%
,t\right) \ 1^{\left[ 2\right] }\qquad ,  \label{I24}
\end{equation}
where $1^{\left[ 2\right] }$ is the unit second order tensor, $\left[ ...%
\right] $\ stands for the tensorial product of vectors,\ and it can be
noticed that
\begin{equation}
P^{\left[ 2\right] }\left( \mathbf{r},t\right) =m\ I_{n}^{\left[ 2\right]
}\left( \mathbf{r},t\right) -m\ n\left( \mathbf{r},t\right) \left[ \mathbf{v}%
\left( \mathbf{r},t\right) \mathbf{v}\left( \mathbf{r},t\right) \right]
\label{I25}
\end{equation}
is the pressure tensor of classical hydrodynamics. Neglecting the terms
quadratic in the drift velocity, combining the above equations \ we obtain
that
\begin{equation}
I_{n}^{\left[ 2\right] }\left( \mathbf{r},t\right) =\xi _{n\alpha }n^{\gamma
_{\alpha }}\left( \mathbf{r},t\right) \qquad ,  \label{I26}
\end{equation}
where
\begin{equation}
\xi _{n\alpha }=\frac{2}{3m}\frac{5}{5-3\alpha }\ \frac{\mathcal{C}_{\alpha
}\left( \mathbf{r},t\right) }{\tilde{\beta}_{\alpha }\left( \mathbf{r}%
,t\right) }\qquad ,  \label{I27}
\end{equation}
is the quantity present in Eq. (\ref{20}).

\renewcommand{\thesection}{Appendix B}%
%

\section{The Grand-Canonical Probability Distribution}

\setcounter{equation}{0}%
%
\renewcommand{\theequation}{B.\arabic{equation}}%
%

We introduce now the Taylor series expansion in \textbf{p} in the Lagrange
multiplier of Eq. (\ref{29}), namely
\begin{equation}
\varphi _{1}\left( \mathbf{r},\mathbf{p};t\right) =\mathcal{F}\left( \mathbf{%
r},t\right) +\mathcal{\mathbf{F}}\left( \mathbf{r},t\right) \cdot \mathbf{p}+%
\mathcal{F}^{\left[ 2\right] }\left( \mathbf{r},t\right) \otimes \left[
\mathbf{pp}\right] +...\qquad ,  \label{II1}
\end{equation}
where $\mathcal{F}=\varphi _{1}$ with $\mathbf{p=0}$, $\left[ \mathcal{%
\mathbf{F}}\right] _{i}=\partial \varphi _{1}/\partial p_{i}$ $\left(
i=x,y,z\right) $ with $\mathbf{p=0}$, $\mathcal{F}^{\left[ 2\right] }$ is
the tensor of components $\mathcal{F}_{ij}=\partial ^{2}\varphi
_{1}/\partial p_{i}\partial p_{j}$ with $\mathbf{p=0}$, and $\left[ \mathbf{%
pp}\right] $ is the tensorial product of the vectors $\mathbf{p}$ and $%
\otimes $ stands for fully contracted tensorial product, etc. Hence, for the
quantity $Z_{GC}\left( t\right) $ which ensures the normalization of the
distribution we do have that
\[
\bar{\varrho}\left( \Gamma \mid t\right) =\frac{1}{Z_{GC}\left( t\right) }%
\exp \left\{ -\int d^{3}r\ d^{3}p\left[ \mathcal{F}_{\alpha }\left( \mathbf{r%
},t\right) +\mathcal{\mathbf{F}}_{\alpha }\left( \mathbf{r},t\right) \cdot
\mathbf{p}+\right. \right.
\]
\begin{equation}
\left. \left. +\mathcal{F}_{\alpha }^{\left[ 2\right] }\left( \mathbf{r}%
,t\right) \otimes \left[ \mathbf{pp}\right] +...\right] \ \hat{n}_{1}\left(
\mathbf{r},\mathbf{p}\mid \Gamma \right) \right\} \qquad .  \label{II2}
\end{equation}

Rewriting the Lagrange multipliers $\mathcal{F}$ as
\begin{equation}
\mathcal{F}\left( \mathbf{r},t\right) =A\left( \mathbf{r},t\right) \qquad ,
\label{II3}
\end{equation}
\begin{equation}
\mathcal{\mathbf{F}}\left( \mathbf{r},t\right) \cdot \mathbf{p}=\mathbf{\nu }%
_{n}\left( \mathbf{r},t\right) \cdot \frac{\mathbf{p}}{m}+\frac{p^{2}}{2m}\
\mathbf{\nu }_{h}\left( \mathbf{r},t\right) \cdot \frac{\mathbf{p}}{m}\qquad
,  \label{II4}
\end{equation}
\begin{equation}
\mathcal{F}^{\left[ r\right] }\left( \mathbf{r},t\right) \otimes \left[
\mathbf{p}...\left( r-times\right) ...\mathbf{p}\right] =F_{n}^{\left[ r%
\right] }\left( \mathbf{r},t\right) \otimes u_{n}^{\left[ r\right] }\left(
\mathbf{p}\right) +F_{h}^{\left[ r\right] }\left( \mathbf{r},t\right)
\otimes u_{n}^{\left[ r\right] }\left( \mathbf{p}\right) \quad ,  \label{II5}
\end{equation}
we recover Eq. (\ref{30}) [we recall that the Lagrange multipliers for
energy and density are not independent but related by the law as shown in
Ref. \cite{LVR03}.

\renewcommand{\thesection}{Appendix C}%
%

\section{Radiation Under Flow}

\setcounter{equation}{0}%
%
\renewcommand{\theequation}{C.\arabic{equation}}%
%

Using the auxiliary distribution of Eq. (\ref{49}) -- which, we recall
produces for the basic variables and only the basic variables, in this case $%
\hat{H}$ and $\mathbf{\hat{I}}_{h}$, the same average value as the proper
nonequilibrium distribution, cf. preceding article --, it follows that
\begin{equation}
E\left( t\right) =Tr\left\{ \sum\limits_{\mathbf{k}}\hslash \omega _{\mathbf{%
k}}\ a_{\mathbf{k}}^{\dagger }a_{\mathbf{k}}\ \bar{\varrho}\left( t,0\right)
\right\} =\sum\limits_{\mathbf{k}}\hslash \omega _{\mathbf{k}}Tr\left\{ a_{%
\mathbf{k}}^{\dagger }a_{\mathbf{k}}\ \bar{\varrho}\left( t,0\right)
\right\} \qquad ,  \label{III1}
\end{equation}
\begin{equation}
N\left( t\right) =Tr\left\{ \sum\limits_{\mathbf{k}}a_{\mathbf{k}}^{\dagger
}a_{\mathbf{k}}\ \bar{\varrho}\left( t,0\right) \right\} =\sum\limits_{%
\mathbf{k}}Tr\left\{ a_{\mathbf{k}}^{\dagger }a_{\mathbf{k}}\ \bar{\varrho}%
\left( t,0\right) \right\} \qquad ,  \label{III2}
\end{equation}
and after calculation
\begin{equation}
Tr\left\{ a_{\mathbf{k}}^{\dagger }a_{\mathbf{k}}\ \bar{\varrho}\left(
t,0\right) \right\} =\nu _{\mathbf{k}}\left( t\right)  \label{III3}
\end{equation}
with $\nu _{\mathbf{k}}\left( t\right) $ given by Eq. (\ref{52}) [a
shifted-Planck population]. Considering the case of a weak flux, we can
write
\begin{equation}
\nu _{\mathbf{k}}\left( t\right) =\left[ \exp \left\{ F_{h}\left( t\right) \
\hslash \omega _{\mathbf{k}}\right\} -1\right] ^{-1}+...  \label{III4}
\end{equation}
where $...$ stands for an expansion of Eq. (\ref{52})\ up to second order in
$\mathbf{\nu }_{h}$. Using Eq. (\ref{III4}) in Eq. (\ref{III1}) and (\ref
{III2}) one obtains the expression of Eqs. (\ref{53}) and (\ref{54}).

Let us now consider the insufficient condition in which the flux is ignored,
and we resort to USM in Renyi's approach, introducing the heterotypical
distribution of Eq. (\ref{59}), is the one of Eq. (\ref{60}) where
\begin{equation}
E_{o}\left( t\right) =\left\langle \hat{H}\right\rangle _{o}=Tr\left\{ \hat{H%
}\bar{\varrho}_{\alpha }\left( t,0\right) \right\} \qquad ,  \label{III5}
\end{equation}
\begin{equation}
\sigma _{e}^{2}\left( t\right) =\left\langle \hat{H}\hat{H}\right\rangle
_{o}-\left\langle \hat{H}\right\rangle _{o}^{2}=Tr\left\{ \left( \hat{H}%
-\left\langle \hat{H}\right\rangle _{o}\right) ^{2}\bar{\varrho}_{\alpha
}\left( t,0\right) \right\} \qquad ,  \label{III6}
\end{equation}
after performing an expansion around $\alpha =1$ in the exponent of the
escort probability (cf. Appendix \textbf{A} in \textbf{I}) and retaining the
first contribution assuming $\alpha $ near $1$; it can be noticed that these
average values are dependent on the infoentropic index $\alpha $. Moreover,
\begin{equation}
E_{o}\left( t\right) =Tr\left\{ \sum\limits_{\mathbf{k}}\hslash \omega _{%
\mathbf{k}}a_{\mathbf{k}}^{\dagger }a_{\mathbf{k}}\bar{\varrho}\left(
t,0\right) \right\} =\sum\limits_{\mathbf{k}}\hslash \omega _{\mathbf{k}%
}Tr\left\{ a_{\mathbf{k}}^{\dagger }a_{\mathbf{k}}\bar{\varrho}_{\alpha
}\left( t,0\right) \right\} =\sum\limits_{\mathbf{k}}\hslash \omega _{%
\mathbf{k}}\mathcal{N}_{\mathbf{k}}\left( t\right) ,  \label{III7}
\end{equation}
where $\mathcal{N}_{\mathbf{k}}\left( t\right) $ is given in Eq. (\ref{62}).%

\renewcommand{\thesection}{Appendix D}%
%

\section{Ideal Gas in a Finite Box}

\setcounter{equation}{0}%
%
\renewcommand{\theequation}{D.\arabic{equation}}%
%

In the conventional approach using the grand-canonical ensemble we do have
the well known results that the average energy is given by
\begin{equation}
E=\sum\limits_{\mathbf{k\sigma }}\epsilon _{\mathbf{k\sigma }}f_{\mathbf{%
k\sigma }}=\int\limits_{0}^{\infty }d\epsilon \ g\left( \epsilon \right) \
\epsilon \ f\left( \epsilon \right) \qquad ,  \label{IV1}
\end{equation}
and the average number of particles by
\begin{equation}
N=\sum\limits_{\mathbf{k\sigma }}f_{\mathbf{k\sigma }}=\int\limits_{0}^{%
\infty }d\epsilon \ g\left( \epsilon \right) \ f\left( \epsilon \right)
\qquad ,  \label{IV2}
\end{equation}
where
\begin{equation}
f\left( \epsilon \right) =\left[ \exp \left\{ \beta \left( \epsilon -\mu
\right) \right\} +1\right] ^{-1}\qquad ,  \label{IV3}
\end{equation}
with $\beta =1/k_{B}T$ and $\mu $ is the chemical potential, and $\epsilon _{%
\mathbf{k\sigma }}=\hslash ^{2}k^{2}/2m$.

Using for density of states $g\left( \epsilon \right) $ the expression of
Eq. (\ref{68}) in Eqs. (\ref{IV1}) and (\ref{IV2}), after taking the
statistically nondegenerate limit, i.e. neglecting in Eq. (\ref{IV3}) $1$ in
comparison with the exponential, there follow the expressions of Eqs. (\ref
{69}) and (\ref{70}) in the text.

We consider now the unconventional approach, introducing the insufficiency
in Fisher's sense of taking for $g\left( \epsilon \right) $ only the first
term on the right of Eq. (\ref{68}) -- meaning the expression corresponding
to the thermodynamic limit --, and resorting to Renyi's approach, that is,
using the statistical operator of Eq. (\ref{72}) and in the calculation of
averages the escort probability in term of it, it follows that the average
value of energy is given by
\begin{equation}
E\simeq \left\langle \hat{H}\right\rangle _{o}-\left( \alpha -1\right) F_{h}%
\left[ \left\langle \hat{H}\hat{H}\right\rangle _{o}-\left\langle \hat{H}%
\right\rangle _{o}^{2}\right] -\left( \alpha -1\right) F_{n}\left[
\left\langle \hat{H}\hat{N}\right\rangle _{o}-\left\langle \hat{H}%
\right\rangle _{o}\left\langle \hat{N}\right\rangle _{o}\right] ,
\label{IV4}
\end{equation}
and for the average value of the particle number
\begin{equation}
N\simeq \left\langle \hat{N}\right\rangle _{o}-\left( \alpha -1\right) F_{n}%
\left[ \left\langle \hat{N}\hat{N}\right\rangle _{o}-\left\langle \hat{N}%
\right\rangle _{o}^{2}\right] -\left( \alpha -1\right) F_{h}\left[
\left\langle \hat{H}\hat{N}\right\rangle _{o}-\left\langle \hat{H}%
\right\rangle _{o}\left\langle \hat{N}\right\rangle _{o}\right] ,
\label{IV5}
\end{equation}
where
\begin{equation}
\left\langle ...\right\rangle =Tr\left\{ ...\bar{\varrho}_{\alpha }\right\}
\qquad ,  \label{IV6}
\end{equation}
and it has been taken $\alpha $ is near $1$, i.e. the case of a finite but
long box.

Performing the calculations, a lengthy but straightforward task, it follows
that
\begin{equation}
E=\frac{3}{2}N_{o}F_{h}^{-1}-\left( \alpha -1\right) \left( \frac{15}{8}%
N_{o}F_{h}^{-1}+\frac{3}{2}N_{o}F_{n}F_{h}^{-1}\right) \qquad ,  \label{IV7}
\end{equation}
\begin{equation}
N=N_{o}-\left( \alpha -1\right) N_{o}\left( \frac{3}{2}+F_{n}\right) \qquad .
\label{IV8}
\end{equation}

It should be noticed that the averages of the kind of Eq. (\ref{IV5})
present in these expression are dependent on $\alpha $. If, in a first
approximation we approximate these average by taking $\alpha \simeq 1$, from
Eq. (\ref{IV7}) and (\ref{IV8}) there follows Eq. (\ref{74}) and
consequently the expression for the infoentropic index $\alpha $ as given by
Eq. (\ref{75}).

\newpage
\bibliographystyle{prsty}
\bibliography{bibliog}

\newpage

\begin{center}
\begin{tabular}[t]{|c|c|c|}
\hline
$L_{QW}\left( \text{\AA }\right) \ $ & $\alpha $ & $\ \ \ \Theta \left(
K\right) \quad $ \\ \hline
$15$ & $\ \ 0.698\quad $ & $48$ \\ \hline
$30$ & $0.714$ & $26$ \\ \hline
$50$ & $0.745$ & $17$ \\ \hline
$80$ & $0.764$ & $10$ \\ \hline
\end{tabular}

Values of the infoentropic index $\alpha $ and kinetic temperature $\Theta $
fitting the luminescence spectrum as shown in Figure $1.$ \newpage %
\includegraphics[width=10cm]{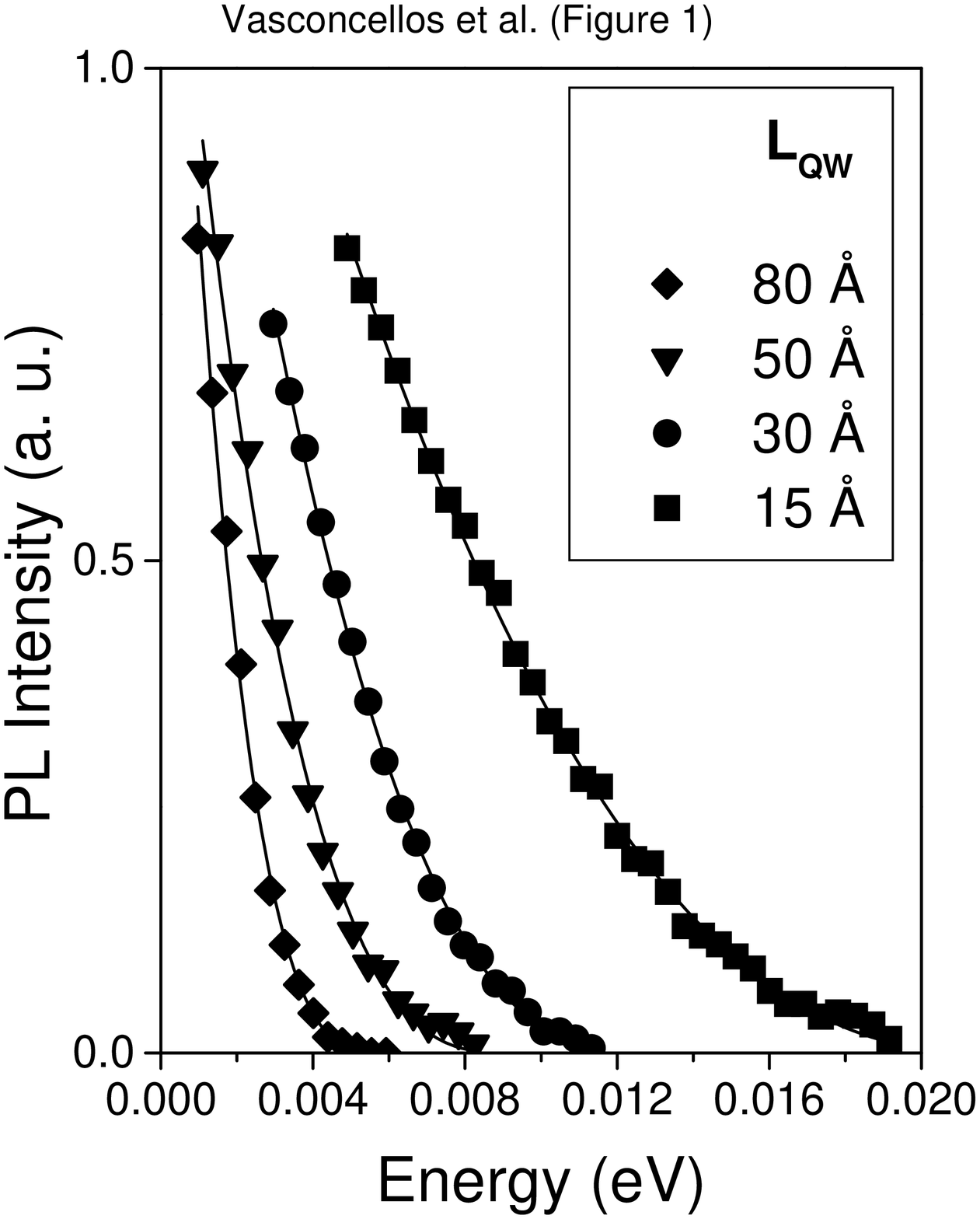} 
\end{center}

\begin{description}
\item[Figure 1:]  Comparison of theory (full line) and experimental results
(filled geometric figures) in the luminescence spectra (high frequency side
or Shockley-Roosbroecke domain purely dependent on the carriers' dynamics),
for several samples under identical processes of growth, but with different
length of the quantum well. The corresponding values of the infoentropic
parameter $\alpha $ and the kinetic temperature $\Theta $ are given in Table
\textbf{I.}
\end{description}

\begin{center}
\newpage \includegraphics[width=10cm]{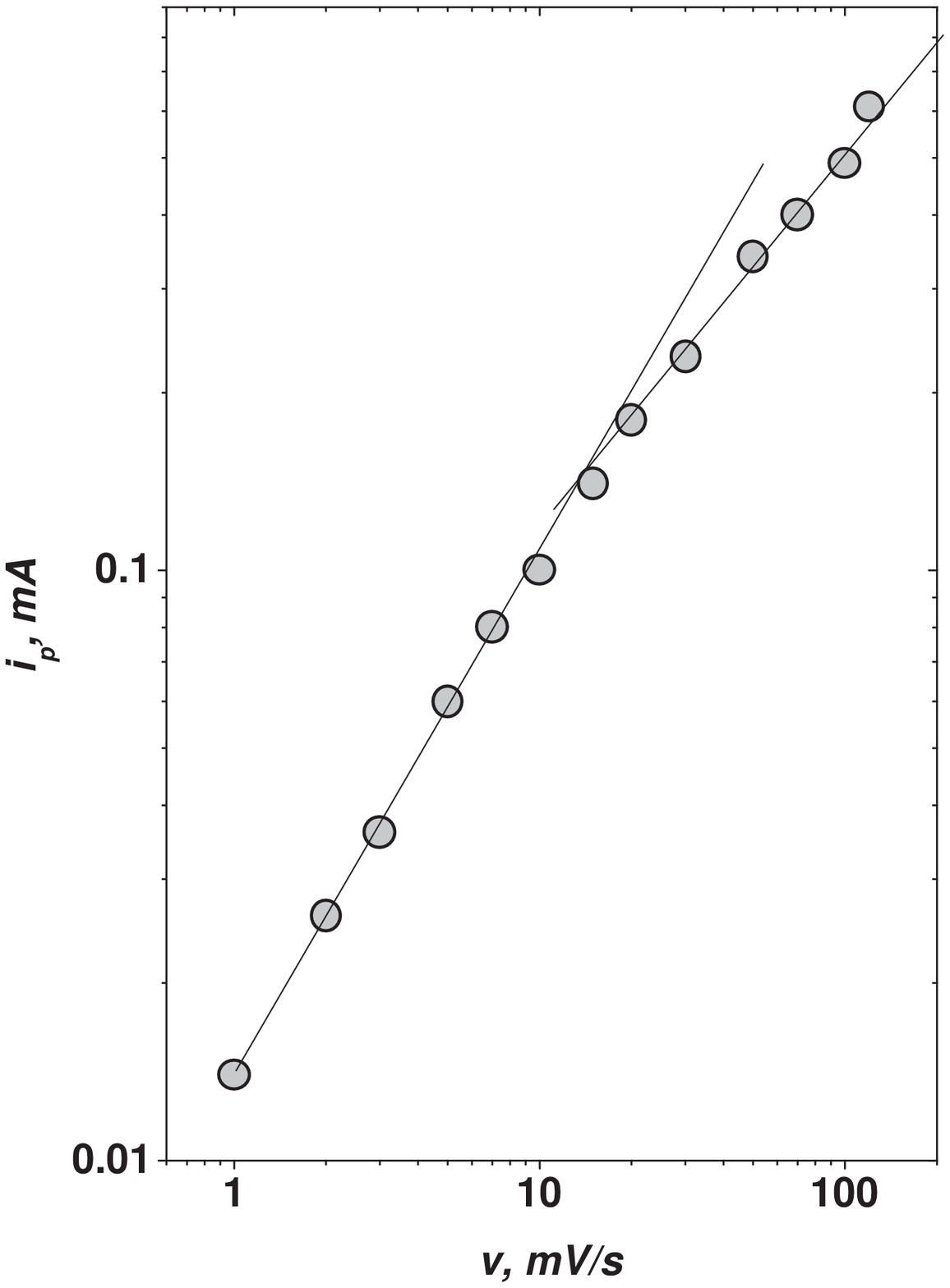}
\end{center}

\begin{description}
\item[Figure 2:]  Logarithmic plot of the current peak value with the
scanning velocity $v$ of the applied electric field, in experiments of
cyclic voltammetry in microbatteries with nanometric thin film fractal-like
electrodes. The tangent at each point gives the value of the index $\xi $ of
Eq. (29); the lower and upper straight lines are approximate fittings in
regions where $\xi $ varies smoothly [cf. Eq. (31)].
\end{description}

\end{document}